\documentclass[english]{elsarticle}
\usepackage[T1]{fontenc}
\usepackage[latin9]{inputenc}
\usepackage{float}
\usepackage{amsmath}
\usepackage{amsthm}
\usepackage{graphicx}

\makeatletter

\newcommand{\lyxdot}{.}

\floatstyle{ruled}
\newfloat{algorithm}{tbp}{loa}
\providecommand{\algorithmname}{Algorithm}
\floatname{algorithm}{\protect\algorithmname}

\usepackage{fullpage}
\usepackage{url}
\usepackage{tabularx}

\makeatother

\usepackage{babel}
\begin{document}
\title{Accelerating the simulation of kinetic shear Alfvén waves with a dynamical low-rank approximation\tnoteref{label1}}
\author[uibk]{Lukas Einkemmer\corref{cor1}} \ead{lukas.einkemmer@uibk.ac.at}
\address[uibk]{University of Innsbruck, Austria}
\cortext[cor1]{Corresponding author}
\begin{abstract} We propose a dynamical low-rank algorithm for a gyrokinetic model that is used to describe strongly magnetized plasmas. The low-rank approximation is based on a decomposition into variables parallel and perpendicular to the magnetic field, as suggested by the physics of the underlying problem. We show that the resulting scheme exactly recovers the dispersion relation even with rank $1$. We then perform a simulation of kinetic shear Alfvén waves and show that using the proposed dynamical low-rank algorithm a drastic reduction (multiple orders of magnitude) in both computational time and memory consumption can be achieved. We also compare the performance of robust first and second-order projector splitting, BUG (also called unconventional), and augmented BUG integrators as well as a FFT-based spectral and Lax--Wendroff discretization.
\end{abstract}
\maketitle

\section{Introduction}

Nuclear fusion promises clean, abundant, and cheap energy. While significant
progress has been mode over the last half century, there are still
many challenges that remain (e.g.~control of plasma instabilities
and understanding turbulence and associated transport phenomena).
Numerical simulations are one of the primary ways to gain insight
into the behavior of such systems and to inform design decisions for
future fusion devices.

However, many phenomena that are of interest (particularly in the
edge and the scrape-off layer) require a kinetic description. Kinetic
models are posed in a relatively high dimensional phase space (four
dimensions for drift-kinetic type models, up to five dimensions for
gyrokinetic models, and six dimensions for the full Vlasov equation).
If this phase space is discretized directly the computational cost
scales as $\mathcal{O}(n^{d})$, where $n$ is the number of grid
points per dimension and $d$ is the number of dimensions. This exponential
increase of storage and computational cost is often referred to as
the curse of dimensionality. It makes realistic simulations often
prohibitively expensive. Thus, most commonly particle in cell (PIC)
methods are employed in large scale codes (see, e.g., \citep{Verboncoeur2005}).
In this approach only the field quantities (which are at most three
dimensional) are discretized and the kinetic behavior is resolved
by particles dynamics. This, for some problems, can drastically reduce
the computational effort needed, especially if high-performance computing
systems are used. However, particle in cell methods suffer from numerical
noise and slow convergence (as the square root in the number of particles).
They are still drastically more expensive than fluid or magnetohydrodynamics
simulations.

More recently, dynamical low-rank methods have been proposed in \citep{Einkemmer2018}
to solve kinetic problems. These methods originate from early work
in quantum mechanics \citep{Meyer2009,Lubich2008} and were later
considered in a mathematical framework for for ordinary differential
equations (see, e.g., \citep{Koch2007,Nonnenmacher2008}). In the
latter context, many advances such as robust integrators \citep{Lubich2014,Ceruti2022},
rank adaptive methods \citep{Ceruti2022a,Ceruti2022b,Hochbruck2023,Hauck2022},
and generalization to various tensor formats \citep{Lubich2013,Lubich2015,Lubich2018,Ehrlacher2017,Ceruti2020a,Ceruti2022b}
have been made. While such methods can be applied in a rather generic
way to ordinary or partial differential equation, an efficient algorithm
is only obtained if a suitable decomposition of variables is chosen
that allows us to run the simulation with a small to moderate rank.
For kinetic problems primarily a decomposition between spatial and
velocity variables has been performed (see \citep{Einkemmer2019,Einkemmer2020,Einkemmer2021,Peng2020,Peng2021,Ding2021,Coughlin2022,Kusch2023};
some work on tensor decomposition also exists \citep{Kormann2015,Einkemmer2018,AllmannRahn2022,Guo2022a}).
It turns out that for kinetic problems this has a number of advantages.
In particular, for collisional problems the associated fluid limit
is often low-rank \citep{Ding2021,Einkemmer2021d} as are various
linearized models of the Vlasov equation (e.g.~Landau damping is
a low-rank phenomenon \citep{Einkemmer2020}). Because of this, for
many problems, dynamical low-rank approximation perform very well
and drastically reduce the computational effort required. In some
situations full six-dimensional kinetic simulations can be performed
using a desktop computer \citep{Cassini2021}. One shortcoming of
low-rank methods in the context of kinetic equation is that they are
not conservative. However, recently, significant progress has been
made to make such methods mass, momentum, and energy conservative
\citep{Einkemmer2021a,Guo2022,Einkemmer2022}. 

Almost all of the previous work has been done in the context of the
Vlasov--Poisson or Vlasov--Maxwell equations. For strongly magnetized
plasmas (such as those found in fusion plasmas) using the Vlasov equation
directly implies that we need to resolve the extremely fast cyclotron
frequency. Thus, in this case analytically reduced models, called
gyrokinetics, are most commonly used. While those retain the general
structure of the Vlasov equation there are important differences.
In this paper, our goal is to make the first step towards using dynamical
low-rank simulations for strongly magnetized plasmas. A major change
in this case is how the low-rank decomposition is done. Since in strongly
magnetized plasmas the dynamics parallel and the perpendicular to
the magnetic field often only weakly couples (this was already exploited
in the Hasegawa--Wakatani standard model \citep{Hasegawa1983}),
we decompose into variables perpendicular and parallel to the magnetic
field. This, precisely because it respects the physics of the underlying
problem, endows the proposed dynamical low-rank approximation with
a number of advantageous properties, as we will show.

In this paper, we look at a gyrokinetic model with straight field
lines that can be used to study kinetic shear Alfvén waves. After
introducing the gyrokinetic model that we use (section \ref{sec:model-equation})
we will
\begin{itemize}
\item discuss in section \ref{sec:DLR-algorithm} the chosen decomposition
and derive the equations of motion for the dynamical low-rank approximation.
A particularly important point here is the treatment of the vector
potential. We will also discuss the time and space discretization
used;
\item show in section \ref{sec:properties-dlr-algo} that the proposed dynamical
low-rank algorithm when applied to the kinetic Alfvén wave problem
exactly reproduces the dispersion relation of the original problem.
This is an important result as it shows us that, within the validity
of linear theory, rank 1 is sufficient in order to obtain the correct
behavior. It also allows us to understand why dynamical low-rank approaches
work well in this case and justifies the decomposition chosen;
\item in section \ref{sec:numerical-results} we present numerical simulation
for kinetic Alfvén waves and compare them to analytic theory as well
as a semi-Lagrangian simulation of the full problem. We observe excellent
agreement and a drastic reduction in storage and computational cost.
In section \ref{sec:comparison} we then compare different robust
integrators and spatial discretization strategies as well as their
respective computational cost.
\end{itemize}

\section{Model equation\label{sec:model-equation}}

In this paper we will assume that the magnetic field lines are straight
and point in the $z$-direction. This significantly simplifies the
equations of gyrokinetics since the terms that describe the curvature
of the magnetic field vanish. In non-dimensional form we obtain an
evolution equation for the four-dimensional electron particle density
function $f(t,x,y,z,v)$
\begin{align}
\partial_{t}f+v\partial_{z}f+\frac{1}{M_{e}}\left(\partial_{z}\phi+\partial_{t}A\right)\partial_{v}f & =0,\label{eq:drift-kinetic}
\end{align}
where $M_{e}$ is the ion to electron mass ratio and $v$ is the velocity
in the $z$-direction (no velocity component perpendicular to the
magnetic field is taken into account here). Note that the equations
have been non-dimensionalized with respect to the ion thermal speed
$v_{th,i}$ and the length of the device $L$. This evolution equation
couples to the $z$-component of the electric field $E_{3}=-\partial_{z}\phi-\partial_{t}A$.
Here the scalar potential is denoted by $\phi(t,x,y,z)$ and the $z$
component of the vector potential is denoted by $A(t,x,y,z)$. The
potentials are self-consistently determined by 
\begin{equation}
(\partial_{xx}+\partial_{yy})\phi=C_{P}\rho\qquad\text{and}\qquad(\partial_{xx}+\partial_{yy})A=C_{A}j,\label{eq:field-equations}
\end{equation}
where the charge density and current are given by
\[
\rho=1-\int f\,dv,\qquad\qquad j=-\int vf\,dv
\]
and 
\[
C_{P}=\frac{1}{(\rho_{i}/L)^{2}},\qquad\qquad C_{A}=\frac{\beta}{(\rho_{i}/L)^{2}}.
\]

The plasma $\beta$ (the ratio of plasma pressure to magnetic pressure)
and $\rho_{i}/L$ (the ratio between the ion gyroradius and the size
of the device) are dimensionless parameters. For more details on this
model we refer to \citep{Dannert2004,Lu2021} and the gyrokinetic
literature in general.

We note that although the structure of this equation is similar to
the Vlasov--Poisson or Vlasov--Maxwell equations, there are important
differences. The equation to determine the potential $\phi$ only
has the part of the Laplacian that is perpendicular to the magnetic
field (i.e.~$\Delta_{\perp}=\Delta-\partial_{zz}=\partial_{xx}+\partial_{yy}$).
Thus, for each point in $z$ we solve independent 2D dimensional Poisson
problems. The equation for the vector potential is similar. However,
in the evolution equation (\ref{eq:drift-kinetic}) we actually need
the time derivative, i.e.~$\partial_{t}A$, and not the vector potential
directly. This has some significant ramifications for the numerical
schemes as well as the low-rank approximation proposed here.

\section{Dynamical low-rank algorithm\label{sec:DLR-algorithm}}

The goal of this section is to detail the complexity reduction approach
employed in this paper. Every complexity reduction approach for a
dynamic system requires two ingredients; namely, an approximation
of the high-dimensional quantities and a way to efficiently update
them as the simulation progresses from one time step to the next. 

We employ a low-rank approximation. In this approach the independent
variables $\xi=(x,y,z,v)$ are separated into two groups $\xi=(\xi_{1},\xi_{2})$
and only functions that depend on either $\xi_{1}$ or $\xi_{2}$
are admitted in the low-rank representation. Such an approach has
been shown to be effective for both kinetic equations that appear
in plasma physics (such as the Vlasov--Poisson \citep{Einkemmer2018,Guo2022a}
or Vlasov--Maxwell \citep{Einkemmer2020} equations) and in a number
of transport problems (e.g. radiation transport problems \citep{Peng2020,Ding2021,Peng2021,Kusch2021a}
or the Boltzmann--BGK equation \citep{Einkemmer2019,Einkemmer2021d,Coughlin2022}).
In all of these cases $\xi_{1}$ contains the spatial variables and
$\xi_{2}$ contains all the velocity variables. This reduces a six
(or five) dimensional problem to a three-dimensional problem (although
tensor approximations that further reduce the dimensionality of the
problem have also been proposed \citep{Kormann2015,Einkemmer2018,AllmannRahn2022,Guo2022a}).
Besides numerical evidence, there are also physical reasons why such
an approximation is effective. For example, we know that a low-rank
structure exists in many important solutions to those equations; e.g.
the collisional limit (see \citep{Ding2021,Einkemmer2021,Einkemmer2021d})
and physical phenomena that are described well by a linearization
of the Vlasov equation (e.g.~Landau damping \citep{Einkemmer2020}
or the onset of the two-stream instability \citep{Cassini2021}) have
a low-rank structure. Let us emphasize that, in general, the way this
decomposition is done depends on the problem under consideration.
For example, if a low-rank approximation is used e.g.~in quantum
mechanics \citep{Meyer2009,Lubich2008} or for the chemical master
equation in biology \citep{Jahnke2008,Prugger2023} the choice of
the decomposition (and potentially also the choice of the tensor format)
is very different.

In the present case we deal with plasmas that are strongly magnetized
by an external magnetic field along the $z$-axis. The gyrokinetic
approach removes the fastest scales associated with the cyclotron
frequency, which leads to equations (\ref{eq:drift-kinetic}) and
(\ref{eq:field-equations}). Those equations have significant differences
compared to the classic electrostatic Vlasov--Poisson system. This
is seen both in the field equations, where only $\Delta_{\perp}=\partial_{xx}+\partial_{yy}$,
i.e.~the Laplacian perpendicular to the magnetic field appears, as
well as in the evolution equation, where kinetic effects are only
taken into account in the $z$-direction. The solution of such problems
commonly admits a structure where the degrees of freedom associated
with the $z$-direction can be separated from the perpendicular directions.
This is not only the case for kinetic problems but is, e.g., also
used in the Hasegawa--Wakatani standard model \citep{Hasegawa1983}.
Thus, we consider here a decomposition with $\xi_{1}=(x,y)$ and $\xi_{2}=(z,v).$
The particle density function is then represented as
\begin{equation}
f(t,x,y,z,v)=\sum_{i=1}^{r}\sum_{j=1}^{r}X_{i}^{f}(t,x,y)S_{ij}^{f}(t)V_{j}^{f}(t,z,v),\label{eq:lr-f}
\end{equation}
where $X_{i}^{f}$, $S_{ij}^{f}$ , and $V_{j}^{f}$ are the corresponding
low-rank factors (which depend only on $(x,y)$ or $(z,v)$ and are
thus at most two dimensional). We assume the orthogonality conditions
\[
\langle X_{i}^{f},X_{j}^{f}\rangle_{xv}=\delta_{ij},\qquad\qquad\text{and}\qquad\qquad\langle V_{i}^{f},V_{j}^{f}\rangle_{zv}=\delta_{ij},
\]
where $\langle f,g\rangle_{xy}=\int fg\,d(x,y)$ and $\langle f,g\rangle_{zv}=\int fg\,d(z,v)$
are the usual $L^{2}$ inner products. The rank of the approximation
is denoted by $r$. Storing the full particle density would require
$\mathcal{O}(n^{4})$, where $n$ is the number of grid points in
each of the spatial and velocity directions, for simplicity here assumed
to be equal. The low-rank approximation reduces this to $\mathcal{O}(rn^{2})$
. Increasing the rank increases the accuracy of the method. However,
it also increases the storage and computational cost to run the simulation.
We can also see from this discussion that using this decomposition,
in the present case, is more appropriate than using $\xi_{1}=(x,y,z)$
and $\xi_{2}=(v)$, as this would still yield a three dimensional
problem once the low-rank approximation has been performed and thus
increase computational cost.

A major difference compared to the commonly used dynamical low-rank
approximations for kinetic equations in the literature is that the
field quantities are also subject to a low-rank approximation. That
is, we consider
\begin{equation}
\phi(t,x,y,z)=\sum_{ij}X_{i}^{\phi}(t,x,y)S_{ij}^{\phi}(t)V_{j}^{\phi}(t,z),\qquad\text{and}\qquad\partial_{t}A(t,x,y,z)=\sum_{ij}X_{i}^{\partial_{t}A}(t,x,y)S_{ij}^{\partial_{t}A}(t)V_{j}^{\partial_{t}A}(t,z).\label{eq:lr-phiA}
\end{equation}
Note that in the algorithm we only use $\partial_{t}A$. The vector
potential $A$ is only used to compute the magnetic energy but otherwise
does not need to be stored. From now on we will assume, if not indicated
otherwise, that all summations run from $1$ to $r$. 

The second ingredient required in a complexity reduction scheme is
a way to propagate the solution forward in time. For low-rank approximation
there are two choices. First, one can start with a low-rank approximation
and conduct a time step. This will increase the rank, by how much
depends on the problem under consideration and the numerical method
used. Then a truncation step is performed in order to reduce the rank
to acceptable levels, see e.g.~\citep{Kormann2015,Guo2022a}. This
is often called the step-truncation approach. It should be emphasized
that since the field quantities are also subject to a low-rank approximation,
the situation here is quite different from the Vlasov--Poisson equation
(for which the step-truncation approach has been considered in the
references mentioned above). In our case a term like $\partial_{z}\phi\partial_{v}f$
has, in general, rank $r^{2}$, while for the Vlasov--Poisson equation
the electric field is not subject to a low-rank approximation and
thus such a term is only of rank $r$. This, obviously, increases
the memory required (in the intermediate steps) of a step-truncation
approach.

We will use here the more recently developed dynamical low-rank approach
\citep{Koch2007,Einkemmer2018}. In this approach equations of motions
for the low-rank factors are derived in the continuous setting which
are then discretized (as any other PDE) using a numerical method.
This has two primary advantages. First, the rank of the solution never
increases and thus only approximations of rank $r$ have to be stored.
Second, it allows us to derive the low-rank approximation on the continuous
level. This yields a set of PDEs to which we can then apply an appropriate
numerical method. While we will here mainly use explicit integrators,
for the equations considered implicit approaches can be interesting
\citep{Lu2021}. Using an implicit scheme, e.g., is very hard in the
step-truncation approach.

In the following we will consider the different parts of equations
of motion separately. In section \ref{subsec:evolution} we will consider
the evolution equation (\ref{eq:drift-kinetic}). This is followed
by the scalar potential $\phi$ in section \ref{subsec:phi} and the
vector potential $\partial_{t}A$ in section \ref{subsec:dtA}. The
details of the robust dynamical low-rank integrators are then discussed
in section \ref{subsec:dlr-algorithm}. Finally, we will consider
the numerical discretization of the derived PDEs for the low-rank
factors (section \ref{subsec:discretization}) and an adaptive time
stepping algorithm (section \ref{subsec:adaptive-step-size}).

\subsection{Evolution equation\label{subsec:evolution}}

Applying the dynamical low-rank approach as described in \citep{Koch2007,Einkemmer2018}
results in a set of evolution equations that couple the low-rank factors
$X_{i}^{f}$, $S_{ij}^{f}$, and $V_{j}^{f}$. In principle that system
of equations can be integrated forward in time. However, it requires
the computation of $(S^{f})^{-1}$. For a sufficiently accurate approximation
or even an overapproximation (i.e.~where the rank is chosen larger
than required) $S^{f}$ has singular values which are close to zero
and thus computing the inverse is ill-conditioned. Although, this
problem has been known for a long time, the first robust dynamical
low-rank integrator which does not suffer from this problem has been
introduced relatively recently \citep{Lubich2014}. This projector-splitting
integrator splits the projector onto the approximation space into
three parts that correspond to the three low-rank factors. By doing
so the three equations decouple which allows us to get rid of the
ill-conditioned inverse (for a detailed discussion in the context
of kinetic equations we refer to \citep{Einkemmer2018}). More recently,
the basis updating Galerkin (BUG) integrator has been proposed in
\citep{Ceruti2022} and improved in \citep{Ceruti2022a}. The idea
is different, but the BUG integrator still uses the same three evolution
equations to perform the update. We will consider both integrators
in the present work. How they combine the three equations outlined
here in order to obtain an approximation to the dynamics is explained
in section \ref{subsec:dlr-algorithm}.

\emph{Equation for K: }In the K step we consider not $X_{i}$ by itself,
but instead
\[
K_{j}=\sum_{i}X_{i}S_{ij}.
\]
This is a major ingredient that allows us to obtain a robust scheme.
Note that we can always obtain $X^{f}$ and $S^{f}$ from $K^{f}$
in a robust way by performing a QR decomposition. In order to obtain
an evolution equation for $K^{f}$ we project the dynamics of the
equation onto the basis in $\xi_{2}=(z,v)$ (i.e. onto the space spanned
by the $V_{j}^{f}$). That is, we have
\[
\partial_{t}K_{j}^{f}=\langle V_{j}^{f},\text{RHS}\rangle_{zv},
\]
where $\text{RHS}$ is the right-hand side of equation (\ref{eq:drift-kinetic}).
Thus, we have
\[
\partial_{t}K_{j}^{f}=\langle V_{j}^{f},-v\partial_{z}f-\frac{1}{M_{e}}(\partial_{z}\phi+\partial_{t}A)\partial_{v}f\rangle_{zv}.
\]
Plugging in the low-rank representations for $f$, equation (\ref{eq:lr-f}),
$\phi$, and $\partial_{t}A$, equation (\ref{eq:lr-phiA}), we get
\begin{align}
\partial_{t}K_{j}^{f} & =-\sum_{l}\langle vV_{j}^{f},\partial_{z}V_{l}^{f}\rangle_{zv}K_{l}^{f}-\frac{1}{M_{e}}\sum_{ln}\left(K_{n}^{\phi}\langle(\partial_{z}V_{n}^{\phi})V_{j}^{f},\partial_{v}V_{l}^{f}\rangle_{zv}+K_{n}^{\partial_{t}A}\langle V_{n}^{\partial_{t}A}V_{j}^{f},\partial_{v}V_{l}^{f}\rangle_{zv}\right)K_{l}^{f}\nonumber \\
 & =-\sum_{l}c_{jl}^{1}K_{l}^{f}-\frac{1}{M_{e}}\sum_{ln}c_{jln}^{2}K_{n}^{\phi}K_{l}^{f}-\frac{1}{M_{e}}\sum_{ln}c_{jln}^{3}K_{n}^{\partial_{t}A}K_{l}^{f},\label{eq:K}
\end{align}
where the coefficients are given by
\[
c_{jl}^{1}=\langle vV_{j}^{f},\partial_{z}V_{l}^{f}\rangle_{zv},\qquad c_{jln}^{2}=\langle V_{j}^{f}\partial_{v}V_{l}^{f},(\partial_{z}V_{n}^{\phi})\rangle_{zv},\qquad c_{jln}^{3}=\langle V_{j}^{f}\partial_{v}V_{l}^{f},V_{n}^{\partial_{t}A}\rangle_{zv}.
\]

Note that the coefficients are constant in time as only $K^{f}$ (and
thus $X^{f}$ and $S^{f}$) are updated during this step. Thus, the
coefficients $c^{1}$, $c^{2}$, and $c^{3}$ can be computed before
performing the time stepping for equation (\ref{eq:K}). Computing
the coefficients is, however, a significant fraction of the overall
run time of the algorithm. This is due to the fact that we have to
compute multiple integrals of the basis functions and the scaling
is not linear in $r$. For example, to store $c_{jln}^{2}$ requires
only $\mathcal{O}(r^{3})$ memory, but computing it has a cost of
$\mathcal{O}(r^{3}n^{2})$. Let us also emphasize that the equation
for $K$ is an ordinary differential equation. In particular, no spatial
derivatives occur and thus this equation is easy to solve and parallelize.

\emph{Equation for S: }In the S step we project onto both the space
spanned by the $X_{i}^{f}$ and the space spanned by the $V_{j}^{f}$.
We have
\begin{align*}
\partial_{t}S_{ij}^{f} & =\langle X_{i}^{f}V_{j}^{f},\text{RHS}\rangle=-\langle X_{i}^{f}V_{j}^{f},v\partial_{z}f+\frac{1}{M_{e}}(\partial_{z}\phi+\partial_{t}A)\partial_{v}f\rangle_{xyzv}.
\end{align*}
Plugging in the low-rank representations for $f$, equation (\ref{eq:lr-f}),
$\phi$, and $\partial_{t}A$, equation (\ref{subsec:dtA}), and using
the orthogonality relation between the basis functions, we get
\begin{align}
\partial_{t}S_{ij}^{f} & =-\sum_{l}S_{il}^{f}\langle vV_{j}^{f},\partial_{z}V_{l}^{f}\rangle_{xyzv}-\frac{1}{M_{e}}\sum_{mnkl}S_{kl}^{f}\langle X_{i}^{f},X_{m}^{\phi}X_{k}^{f}\rangle_{xy}S_{mn}^{\phi}\langle(\partial_{z}V_{n}^{\phi})V_{j}^{f},\partial_{v}V_{l}^{f}\rangle_{zv}\nonumber \\
 & \qquad\qquad-\frac{1}{M_{e}}\sum_{mnkl}S_{kl}^{f}\langle X_{i}^{f},X_{m}^{\partial_{t}A}X_{k}^{f}\rangle_{xy}S_{mn}^{\partial_{t}A}\langle V_{n}^{\partial_{t}A}V_{j}^{f},\partial_{v}V_{l}^{f}\rangle_{zv}\nonumber \\
 & =-\sum_{l}S_{il}^{f}c_{jl}^{1}-\frac{1}{M_{e}}\sum_{nk}\left(\sum_{m}d_{ikm}^{2}S_{mn}^{\phi}\right)\left(\sum_{l}S_{kl}^{f}c_{jln}^{2}\right)-\frac{1}{M_{e}}\sum_{nk}\left(\sum_{m}d_{ikm}^{3}S_{mn}^{\partial_{t}A}\right)\left(\sum_{l}S_{kl}^{f}c_{jln}^{3}\right),\label{eq:S}
\end{align}
where
\[
d_{ikm}^{2}=\langle X_{i}^{f},X_{k}^{f}X_{m}^{\phi}\rangle_{xy},\qquad\qquad d_{ikm}^{3}=\langle X_{i}^{f},X_{k}^{f}X_{m}^{\partial_{t}A}\rangle_{xy}.
\]

Once again this is an ordinary differential equation with coefficients
that are constant in time during that step. Since $S$ has only $r^{2}$
entries, solving equation (\ref{eq:S}) is only a negligible part
of the overall computational cost of the algorithm.

\emph{Equation for L:} In the L step we consider
\[
L_{i}=\sum_{j}S_{ij}V_{j}.
\]
Once again we can easily obtain $S^{f}$ and $V^{f}$ from $L^{f}$
by performing a QR decomposition. We proceed in the same way as before,
except that we now project onto the basis in $\xi_{1}=(x,y)$, i.e.~onto
the space spanned by the $X_{i}^{f}.$ We have
\[
\partial_{t}L_{i}^{f}=\langle X_{i}^{f},\text{RHS}\rangle_{xy}.
\]
 Plugging in the low-rank representations for $f$, equation (\ref{eq:lr-f}),
$\phi$, and $\partial_{t}A$, equation (\ref{eq:lr-phiA}), and using
the orthogonality relation between the basis functions, we get
\begin{align}
\partial_{t}L_{i}^{f} & =\langle X_{i}^{f},-v\partial_{z}f-\frac{1}{M_{e}}(\partial_{z}\phi+\partial_{t}A)\partial_{v}f\rangle_{xy}\nonumber \\
 & =-v\partial_{z}L_{i}^{f}-\frac{1}{M_{e}}\sum_{mk}\langle X_{i}^{f},X_{m}^{\phi}X_{k}^{f}\rangle_{xv}(\partial_{z}L_{m}^{\phi})\partial_{v}L_{k}^{f}-\frac{1}{M_{e}}\sum_{mk}\langle X_{i}^{f},X_{m}^{\partial_{t}A}X_{k}^{f}\rangle_{xv}L_{m}^{\partial_{t}A}\partial_{v}L_{k}^{f}\nonumber \\
 & =-v\partial_{z}L_{i}^{f}-\frac{1}{M_{e}}\sum_{k}(e_{ik}+e_{ik}^{A})\partial_{v}L_{k}^{f},\label{eq:L}
\end{align}
where
\[
e_{ik}(z)=\sum_{m}d_{ikm}^{2}\partial_{z}L_{m}^{\phi}(z),\qquad\qquad e_{ik}^{A}(z)=\sum_{m}d_{ikm}^{3}L_{m}^{\partial_{t}A}(z).
\]

Note that we can also use $e_{ik}=\langle X_{i}^{f}(\partial_{z}\phi)X_{k}^{f}\rangle_{xy}$
and $e_{ik}^{A}=\langle X_{i}^{f}(\partial_{t}A)X_{k}^{f}\rangle_{xy}$.
However, computing these quantities by using $d^{2}$ and $d^{3}$
(which for both the projector splitting and unconventional integrator
can be reused from a previous step in the algorithm) is more efficient.

Note that the free streaming term is treated exactly by the low-rank
approximation. That is, no approximation is made to this term in equation
(\ref{eq:L}). This makes sense as this is an advection in $z$ with
a speed that only depends on $v$ and thus this term only depends
on variables in $\xi_{2}=(z,v)$. For the second term in equation
(\ref{eq:L}) the primary approximation made is due to the fact that
we also have to perform a low-rank approximation of the field quantities.
The result is still an advection in $v$, but with a, in general,
approximated velocity owing to the fact that we can not represent
the field quantities exactly.

\subsection{Equation for the scalar potential\label{subsec:phi}}

In order to determine the scalar potential we have to solve the following
Poisson problem for each $z$ coordinate
\[
\left(\partial_{xx}+\partial_{yy}\right)\phi=C_{P}\rho.
\]
Note that due to the separation of the dynamics parallel and perpendicular
to the magnetic field this differential operator only acts in $\xi_{1}=(x,y)$.
This makes it perfectly compatible with the low-rank approximation
that we employ here. Plugging in the low-rank expansion for $\phi$
and $f$ we get

\begin{align*}
\sum_{j}\left((\partial_{xx}+\partial_{yy})K_{j}^{\phi}\right)V_{j}^{\phi} & =C_{P}\left(1-\sum_{j}K_{j}^{f}\langle V_{j}^{f}\rangle_{v}\right),\qquad\qquad\langle f\rangle_{v}=\int f\,dv.
\end{align*}

Note that $\langle V_{j}^{f}\rangle_{v}$ is not an orthonormal basis
($V_{j}^{f}$ is, but since we integrate in $v$ this property is
lost). However, we can easily find a basis for the right-hand side
of the equation by orthonormalizing $\left(1,\langle V_{1}^{f}\rangle_{v},\dots,\langle V_{r}^{f}\rangle_{v}\right)$.
If we call the resulting orthonormal basis $V_{j}^{\rho}$ we have 

\[
\rho=\sum_{j=1}^{r+1}\rho_{j}V_{j}^{\rho},\qquad\qquad\rho_{j}=\langle\rho,V_{j}^{\rho}\rangle_{z}.
\]
This relation is exact since $f$ has a low-rank structure and thus
the same is true for $\rho$. The rank increases by at most $1$ to
$r+1$. We can also explicitly calculate $\rho_{j}$ from the low-rank
representation of $f$. We have
\begin{align*}
\rho_{j} & =\langle\rho,V_{j}^{\rho}\rangle_{z}\\
 & =\langle1,V_{j}^{\rho}\rangle_{z}-\sum_{i}K_{i}^{f}\left\langle V_{j}^{\rho},\langle V_{i}^{f}\rangle_{v}\right\rangle _{z}.
\end{align*}
If a QR decomposition is used to compute (note that the QR decomposition
is formulated here in the continuous setting; once a discretization
has been performed this becomes a classic QR decomposition)
\[
\sum_{l}V_{l}^{\rho}R_{li}=\left(1,\langle V_{1}^{f}\rangle_{v},\dots,\langle V_{r}^{f}\rangle_{v}\right)_{i}
\]
 we have
\[
\rho_{j}=\Vert1\Vert_{z}\delta_{j1}-\sum K_{i}^{f}R_{j,i+1}.
\]

Since we now have an orthonormal basis we can choose $V_{j}^{\phi}=V_{j}^{\rho}$
which implies at once 
\begin{equation}
(\partial_{xx}+\partial_{yy})K_{j}^{\phi}=\rho_{j}.\label{eq:poisson-phi}
\end{equation}

Thus, we have to solve at most $r+1$ two-dimensional Poisson problems
in $(x,y)$. Any Poisson solver an be used to do that. For a fast
Poisson solver that scales linearly in the number of unknowns, the
computational cost is $\mathcal{O}(rn^{2})$. We note that this is
usually significantly lower compared to computing the coefficients
and integrating the evolution equations (the steps discussed in the
previous section). Thus, this step, if implemented in the way suggested,
only adds a negligible amount of compute time to the algorithm.

\subsection{Iterative scheme for $\partial_{t}A$\label{subsec:dtA}}

It is easy to compute the vector potential $A$ in a similar way as
was described for the scalar potential in the previous section. It
also satisfies a Poisson equation in $(x,y)$ with the primary difference
being that the right-hand side is given by the current and not the
charge density. However, in the evolution equation (\ref{eq:drift-kinetic})
we do not require $A$, but $\partial_{t}A$. We could approximate
$\partial_{t}A$ by forward differences which, since $f$ at the next
time point is not known, would result in an implicit scheme that couples
$f$ and $\partial_{t}A$, which is computationally expensive.

However, we can also derive an equation that directly relates $\partial_{t}A$
to quantities of the particle density function at the same point in
time. By taking the time derivative of 
\[
(\partial_{xx}+\partial_{yy})A=C_{A}j
\]
we get
\begin{equation}
(\partial_{xx}+\partial_{yy})\partial_{t}A=C_{A}\partial_{t}j.\label{eq:eqdtAj}
\end{equation}
We can now express $\partial_{t}j$ using equation (\ref{eq:drift-kinetic})
to get
\[
\partial_{t}j=-\int v\partial_{t}f\,dv=\partial_{z}\int v^{2}f\,dv-\frac{1}{M_{e}}\left(\partial_{z}\phi+\partial_{t}A\right)(1-\rho).
\]

Rearranging equation (\ref{eq:eqdtAj}) we get
\begin{equation}
\left(\partial_{xx}+\partial_{yy}+\frac{C_{A}}{M_{e}}(1-\rho)\right)\partial_{t}A=C_{A}\partial_{z}\int v^{2}f\,dv-\frac{C_{A}}{M_{e}}(\partial_{z}\phi)(1-\rho).\label{eq:poisson-dtA}
\end{equation}

This is still a symmetric operator. Thus, we can apply a conjugate
gradient method to find the solution. Conjugate gradient only requires
the evaluation of the operator applied to an arbitrary vector. For
the differential part this can be done easily as
\[
(\partial_{xx}+\partial_{yy})\partial_{t}A=\sum_{ij}(\partial_{xx}+\partial_{yy})X_{i}^{\partial_{t}A}S_{ij}^{\partial_{t}A}V_{j}^{\partial_{t}A}.
\]

However, the addition or multiplication of two rank $r$ functions
is, in general, not a rank $r$ functions. Thus, we need to multiply
$\partial_{t}A$ and $1-\rho$ together and then add it to the result
of applying the differential operator, which increases the rank to
at most $r+r^{2}$. Then we perform a truncation back to rank $r$
using a singular value decomposition. Since the singular value decomposition
only needs to be performed for $S$ this can be done very efficiently.
Nevertheless, since at least a couple of iterations are required for
convergence, this procedure has to be done multiple times in each
time step and thus computing $\partial_{t}A$ can still be a significant
fraction of the overall run time of the algorithm. We note, however,
that $\partial_{t}A$ does not depend on $v$ and thus the additional
memory required to store intermediate results is much lower than for
a step-truncation algorithm (where the same would need to be done
for $f$). This is, in particular, true for the examples we consider
in this paper, where significantly more grid points are required in
the $v$ direction compared to the spatial directions.

We also need to compute the right-hand side of equation (\ref{eq:poisson-dtA}).
We have $\partial_{z}\int v^{2}f\,dv$ which is of rank $r$. The
term $(\partial_{z}\phi)(1-\rho)$ has (at most) rank $\mathcal{O}(r^{2})$
since we multiply two terms of rank $\mathcal{O}(r)$ together. Thus,
in total the right-hand side has (at most) rank $\mathcal{O}(r^{2}+r)$.
We perform those operations and truncate them using a singular value
decomposition (as above). Since this is only done once, the computational
cost is usually only a very small part of the overall algorithm. 

\subsection{Dynamical low-rank algorithm\label{subsec:dlr-algorithm} }

In principle we can solve the coupled equations for the low-rank factors
(\ref{eq:K}), (\ref{eq:S}), and (\ref{eq:L}) of $f$ together with
the equations for the potentials (\ref{eq:poisson-phi}) and (\ref{eq:poisson-dtA})
to obtain the solution. However, in order to do so we have to invert
$S$, since equation (\ref{eq:K}) is formulated in terms of $K$
(and not $X$) and equation $(\ref{eq:L})$ is formulated in terms
of $L$ (and not $V$). This, however, is undesirable as in cases
where the smallest singular value of $S$ is small the inverse of
$S$ is ill-conditioned. If the smallest singular value of $S$ becomes
larger, however, the approximation is usually very inaccurate (as
we neglect singular values that are relatively large and thus important). 

In order to remedy this problem integrators that are robust to the
presence of small singular values have been developed. In particular,
the projector splitting integrator \citep{Lubich2014} and the recently
developed class of basis updating Galerkin (BUG) integrators \citep{Ceruti2022,Ceruti2022a}
(also called unconventional integrators) have this property. The only
difference is how those algorithm combine equations (\ref{eq:K}),
(\ref{eq:S}), and (\ref{eq:L}) in order to update the numerical
solution. In both cases, in order to increase computational efficiency,
we split the field solves from the time update. The first order projector
splitting integrator is shown in Algorithm \ref{alg:projector-lie},
the first order BUG integrator is shown in Algorithm \ref{alg:bug},
and an augmented version of the BUG integrator (as described in \citep{Ceruti2022a})
is shown in Algorithm \ref{alg:augmented-bug}. For more details on
the projector splitting and BUG integrator, specifically in the context
of kinetic problems, we refer the reader to \citep{Einkemmer2018}
and \citep{Kusch2021a}, respectively. We also note that since the
projector splitting integrator is a splitting scheme it can be raised
to higher order (see, e.g., \citep{Einkemmer2018,Cassini2021}). The
primary difficulty here is that it is not sufficient to compute the
(frozen) field quantities only at the beginning of the time step.
However, we know from \citep{Einkemmer2014d,Einkemmer2014c} that
if we have a first order approximation of the field quantities at
the half step (i.e.~at $t^{n+1/2}=t^{n}+\Delta t/2$) then applying
a Strang splitting scheme using those field quantities retains second
order accuracy. We thus first perform a Lie splitting step to obtain
a first order approximation at the half step, from which the field
quantities that are then used in the Strang splitting are determined.
The second order projector splitting integrator based on this idea
is shown in Algorithm \ref{alg:projector-strang}. Let us also note
that there is currently no second order variant of the BUG integrator.

\begin{algorithm}[H]
\textbf{Input: $X_{i}^{n}$, $S_{ij}^{n}$, $V_{j}^{n}$ }such that
$f(t^{n},x,y,z,v)\approx\sum_{ij}X_{i}^{n}(x,y)S_{ij}^{n}V_{j}^{n}(z,v)$

\textbf{Output: $X_{i}^{n+1}$, $S_{ij}^{n+1}$, $V_{j}^{n+1}$ }such
that $f(t^{n+1},x,y,z,v)\approx\sum_{ij}X_{i}^{n+1}(x,y)S_{ij}^{n+1}V_{j}^{n+1}(z,v)$
\begin{enumerate}
\item Solve the Poisson problem (\ref{eq:poisson-phi}) to obtain $K_{j}^{\phi,n}$
and $V_{j}^{\phi,n}$ (i.e.~$\phi^{n}$) from $f^{n}$.
\item Iteratively solve equation (\ref{eq:poisson-dtA}) to obtain $X_{i}^{\partial_{t}A,n}$,
$S_{ij}^{\partial_{t}A,n}$, and $V_{j}^{\partial_{t}A,n}$ (i.e.~$\partial_{t}A^{n}$)
from $f^{n}$ and $\phi^{n}$.
\item (K step) Perform a time step using time step size $\Delta t$ and
equation (\ref{eq:K}) with initial value $\sum_{i}X_{i}^{n}S_{ij}^{n}$
to obtain $K_{j}^{\star}.$
\item Perform a QR decomposition of $K_{j}^{\star}$ to obtain $X_{i}^{n+1}=Q$
and $S_{ij}^{\star}=R$. 
\item (S step) Perform a time step using time step size $-\Delta t$ and
equation (\ref{eq:S}) with initial value $S_{ij}^{\star}$ to obtain
$S_{ij}^{\star\star}$. 
\item (L step) Perform a time step using time step size $\Delta t$ and
equation (\ref{eq:L}) with initial value $\sum_{j}S_{ij}^{\star\star}V_{j}^{n}$
to obtain $L_{i}^{\star}.$
\item Perform a QR decomposition of $L_{i}^{\star}$ to obtain $V_{j}^{n+1}=Q$
and $S_{ij}^{n+1}=R$.
\end{enumerate}
\caption{First order Lie projector splitting integrator with time step size
$\Delta t=t^{n+1}-t^{n}$, where $t^{n}$ is the time at the beginning
of the current step and $t^{n+1}$ is the time at the end of the current
step. Note that the $S$ step is integrated backward in time. \label{alg:projector-lie}}
\end{algorithm}

\begin{algorithm}[H]
\textbf{Input: $X_{i}^{n}$, $S_{ij}^{n}$, $V_{j}^{n}$ }such that
$f(t^{n},x,y,z,v)\approx\sum_{ij}X_{i}^{n}(x,y)S_{ij}^{n}V_{j}^{n}(z,v)$

\textbf{Output: $X_{i}^{n+1}$, $S_{ij}^{n+1}$, $V_{j}^{n+1}$ }such
that $f(t^{n+1},x,y,z,v)\approx\sum_{ij}X_{i}^{n+1}(x,y)S_{ij}^{n+1}V_{j}^{n+1}(z,v)$
\begin{enumerate}
\item Solve the Poisson problem (\ref{eq:poisson-phi}) to obtain $K_{j}^{\phi,n}$
and $V_{j}^{\phi,n}$ (i.e.~$\phi^{n}$) from $f^{n}$.
\item Iteratively solve equation (\ref{eq:poisson-dtA}) to obtain $X_{i}^{\partial_{t}A,n}$,
$S_{ij}^{\partial_{t}A,n}$, and $V_{j}^{\partial_{t}A,n}$ (i.e.~$\partial_{t}A^{n}$)
from $f^{n}$ and $\phi^{n}$.
\item (K step) Perform a time step using time step size $\Delta t$ and
equation (\ref{eq:K}) with initial value $\sum_{i}X_{i}^{n}S_{ij}^{n}$
to obtain $K_{j}^{\star}.$
\item Perform a QR decomposition of $K_{j}^{\star}$ to obtain $X_{i}^{n+1}=Q$
and throw away $R$.
\item (L step) Perform a time step using time step size $\Delta t$ and
equation (\ref{eq:L}) with initial value $\sum_{j}S_{ij}^{n}V_{j}^{n}$
to obtain $L_{i}^{\star}.$
\item Perform a QR decomposition of $L_{i}^{\star}$ t obtain $V_{j}^{n+1}=Q$
and throw away $R$.
\item Compute $M_{ik}=\langle X_{i}^{n+1},X_{k}^{n}\rangle_{xy}$ and $N_{jl}=\langle V_{j}^{n+1},V_{l}^{n}\rangle_{zv}$. 
\item Recompute $c_{jl}^{1}$, $c_{jln}^{2}$, $c_{jln}^{3}$, $d_{ikm}^{2}$,
$d_{ikm}^{3}$ using $X_{i}^{n+1}$ and $V_{j}^{n+1}$. 
\item (S step) Perform a time step using time step size $\Delta t$ and
equation (\ref{eq:S}) with initial value $\sum_{kl}M_{ik}S_{kl}^{n}N_{jl}$
and the coefficients computed in step 8 to obtain $S_{ij}^{n+1}$. 
\end{enumerate}
\caption{BUG integrator with time step size $\Delta t=t^{n+1}-t^{n}$, where
$t^{n}$ is the time at the beginning of the current step and $t^{n+1}$
is the time at the end of the current step. \label{alg:bug}}
\end{algorithm}

\begin{algorithm}[H]
\textbf{Input: $X_{i}^{n}$, $S_{ij}^{n}$, $V_{j}^{n}$ }such that
$f(t^{n},x,y,z,v)\approx\sum_{ij}X_{i}^{n}(x,y)S_{ij}^{n}V_{j}^{n}(z,v)$

\textbf{Output: $X_{i}^{n+1}$, $S_{ij}^{n+1}$, $V_{j}^{n+1}$ }such
that $f(t^{n+1},x,y,z,v)\approx\sum_{ij}X_{i}^{n+1}(x,y)S_{ij}^{n+1}V_{j}^{n+1}(z,v)$
\begin{enumerate}
\item Solve the Poisson problem (\ref{eq:poisson-phi}) to obtain $K_{j}^{\phi,n}$
and $V_{j}^{\phi,n}$ (i.e.~$\phi^{n}$) from $f^{n}$.
\item Iteratively solve equation (\ref{eq:poisson-dtA}) to obtain $X_{i}^{\partial_{t}A,n}$,
$S_{ij}^{\partial_{t}A,n}$, and $V_{j}^{\partial_{t}A,n}$ (i.e.~$\partial_{t}A^{n}$)
from $f^{n}$ and $\phi^{n}$.
\item (K step) Perform a time step using time step size $\Delta t$ and
equation (\ref{eq:K}) with initial value $\sum_{i}X_{i}^{n}S_{ij}^{n}$
to obtain $K_{j}^{\star}.$
\item Perform a QR decomposition of $K_{j}^{\star}$ to obtain $X_{i}^{\star}=Q$
and throw away $R$.
\item (L step) Perform a time step using time step size $\Delta t$ and
equation (\ref{eq:L}) with initial value $\sum_{j}S_{ij}^{n}V_{j}^{n}$
to obtain $L_{i}^{\star}.$
\item Perform a QR decomposition of $L_{i}^{\star}$ to obtain $V_{j}^{\star}=Q$
and throw away $R$.
\item Form a new basis $X^{\star\star}=\text{orthogonalize}([X^{n},\,X^{\star}])$
and $V^{\star\star}=\text{orthogonalize}([V^{n},\,V^{\star}])$.
\item Recompute $c_{jl}^{1}$, $c_{jln}^{2}$, $c_{jln}^{3}$, $d_{ikm}^{2}$,
$d_{ikm}^{3}$ using $X^{\star\star}$ and $V^{\star\star}$.
\item (S step) Perform a time step using time step size $\Delta t$ and
equation (\ref{eq:S}) with initial value 
\[
S_{ij}=\begin{cases}
S_{ij}^{n} & 1\leq i\leq r,\;1\leq j\leq r\\
0 & \text{otherwise}
\end{cases}
\]
 and the coefficients computed in step 8 to obtain $S_{ij}^{\star\star}$.
\item Perform an SVD to obtain $S_{ij}^{\star\star}=\sum_{k}U_{ik}\sigma_{k}H_{kj}$.
\item Truncate the solution back to rank $r$ by keeping only the $r$ most
important basis functions
\[
\left\{ S_{ij}^{n+1}\right\} _{i,j=1}^{r}=\delta_{ij}\sigma_{i},\qquad\left\{ X_{i}^{n+1}\right\} _{i=1}^{r}=\sum_{j}X_{j}^{\star\star}U_{ji},\qquad\left\{ V_{j}^{n+1}\right\} _{i=1}^{r}=\sum_{i}H_{ji}V_{i}^{\star\star}.
\]
\end{enumerate}
\caption{Augmented BUG integrator with time step size $\Delta t=t^{n+1}-t^{n}$,
where $t^{n}$ is the time at the beginning of the current step and
$t^{n+1}$ is the time at the end of the current step. For the orthogonalization
an arbitrary algorithm (e.g.~modified Gram--Schmidt) can be employed.
Note, however, that for the method described we assume that the orthogonalization
algorithm does not change the first $r$ basis vectors (which are
already orthogonal). This makes the computation of of the initial
value for the S step trivial. \label{alg:augmented-bug}}
\end{algorithm}

\begin{algorithm}[H]
\textbf{Input: $X_{i}^{n}$, $S_{ij}^{n}$, $V_{j}^{n}$ }such that
$f(t^{n},x,y,z,v)\approx\sum_{ij}X_{i}^{n}(x,y)S_{ij}^{n}V_{j}^{n}(z,v)$

\textbf{Output: $X_{i}^{n+1}$, $S_{ij}^{n+1}$, $V_{j}^{n+1}$ }such
that $f(t^{n+1},x,y,z,v)\approx\sum_{ij}X_{i}^{n+1}(x,y)S_{ij}^{n+1}V_{j}^{n+1}(z,v)$
\begin{enumerate}
\item Apply Algorithm \ref{alg:projector-lie} using time step size $\Delta t/2$
and initial value $X_{i}^{n},S_{ij}^{n},V_{j}^{n}$ to obtain $X_{i}^{n+1/2},S_{ij}^{n+1/2},V_{j}^{n+1/2}$.
\item Solve the Poisson problem (\ref{eq:poisson-phi}) to obtain $K_{i}^{\phi,n+1/2}$
and $V_{j}^{\phi,n+1/2}$ (i.e.~$\phi^{n+1/2}$) from $X_{i}^{n+1/2},S_{ij}^{n+1/2},V_{j}^{n+1/2}$.
\item Iteratively solve equation (\ref{eq:poisson-dtA}) to obtain $X_{i}^{\partial_{t}A,n+1/2}$,
$S_{ij}^{\partial_{t}A,n+1/2}$, and $V_{j}^{\partial_{t}A,n+1/2}$
(i.e.~$\partial_{t}A^{n+1/2}$) from $X_{i}^{n+1/2},S_{ij}^{n+1/2},V_{j}^{n+1/2}$
and $\phi^{n+1/2}$.
\item (L step) Perform a time step using time step size $\Delta t/2$ and
equation (\ref{eq:L}) with initial value $\sum_{j}S_{ij}^{n}V_{j}^{n}$
and $\phi^{n+1/2}$, $\partial_{t}A^{n+1/2}$ to obtain $L_{i}^{\star}.$
\item Perform a QR decomposition of $L_{i}^{\star}$ to obtain $V_{j}^{\star}=Q$
and $S_{ij}^{\star}=R$.
\item (S step) Perform a time step using time step size $-\Delta t/2$ and
equation (\ref{eq:S}) with initial value $S_{ij}^{\star}$ and $\phi^{n+1/2}$,
$\partial_{t}A^{n+1/2}$ to obtain $S_{ij}^{\star\star}$.
\item (K step) Perform a time step using time step size $\Delta t$ and
equation (\ref{eq:K}) with initial value $\sum_{i}X_{i}^{\star}S_{ij}^{\star\star}$
and $\phi^{n+1/2}$, $\partial_{t}A^{n+1/2}$ to obtain $K_{j}^{\star}.$
\item Perform a QR decomposition of $K_{j}^{\star}$ to obtain $X_{i}^{n+1}=Q$
and $S_{ij}^{\star\star\star}=R$.
\item (S step) Perform a time step using time step size $-\Delta t/2$ and
equation (\ref{eq:S}) with initial value $S_{ij}^{\star\star\star}$
and $\phi^{n+1/2}$, $\partial_{t}A^{n+1/2}$ to obtain $S_{ij}^{\star\star\star\star}$.
\item (L step) Perform a time step using time step size $\Delta t/2$ and
equation (\ref{eq:L}) with initial value $\sum_{j}S_{ij}^{\star\star\star\star}V_{j}^{\star}$
and $\phi^{n+1/2}$, $\partial_{t}A^{n+1/2}$ to obtain $L_{i}^{n+1}.$
\item Perform a QR decomposition of $L_{i}^{n+1}$ to obtain $V_{j}^{n+1}=Q$
and $S_{ij}^{n+1}=R$.
\end{enumerate}
\caption{Second order Strang projector splitting integrator with time step
size $\Delta t=t^{n+1}-t^{n}$, where $t^{n}$ is the time at the
beginning of the current step and $t^{n+1}$ is the time at the end
of the current step. Note that the $S$ step is integrated backward
in time. \label{alg:projector-strang}}
\end{algorithm}

A peculiarity of the projector splitting integrator is that it requires
us to conduct the S step backward in time. For hyperbolic problems,
such as the one we consider here, this is not an issue and simply
means that we have to negate the right-hand side of equation (\ref{eq:S}).
For both variants of the BUG integrator this is not necessary. However,
from a computational point of view the BUG integrators are more expensive
due to the need to recompute the coefficients with the new basis and,
in the case of the augmented variant, to perform this computation
and the $S$ step for a basis with larger size ($2r$ instead of $r$).
We will consider this computational aspects in more detail in section
\ref{sec:comparison}.

\subsection{Numerical discretization\label{subsec:discretization}}

So far we have developed the dynamical low-rank approximation in the
continuous setting (the approach introduced in \citep{Einkemmer2018}).
That is, no time or space discretization has been performed yet. This
has a number of advantages. First, it makes the distinction between
the error due to the low-rank approximation and the error due to the
spatial and temporal discretization clear. Second, since we start
from a set of partial differential equations for the low-rank factors
by providing an appropriate space and temporal discretization we can
avoid the stability issues that in some cases arises if the numerical
discretization is performed first \citep{Kusch2023}. Third, the numerical
discretization can be tailored to the partial differential equations
obtained for the low-rank factors. In this section we will detail
the spatial and temporal discretization that we will use in our numerical
simulations.

In order to solve for the scalar potential $\phi$ we employ a fast
Fourier transform (FFT) based Poisson solver to obtain the solution
of equation (\ref{eq:poisson-phi}). In order to compute $\partial_{t}A$
we perform a conjugate gradient iteration to obtain a solution of
equation (\ref{eq:poisson-dtA}). The spatial derivatives are computed
using a FFT based approach.

The K equation (\ref{eq:K}) and S equation (\ref{eq:S}) are ordinary
differential equations and thus require no spatial discretization.
For the time integrator we employ the classic Runge--Kutta method
of order 4. 

The L equation (\ref{eq:L}) is a system of advection equations that
depend on ($z,v)$. We consider two distinct ways to discretize this
system.

\emph{Fourier spectral discretization: }First we discretize $L_{i}^{f}$
on an equidistant grid such that $L_{i;\alpha\beta}^{f}(t)\approx L_{i}^{f}(t,z_{\alpha},v_{\beta})$,
where $(z_{\alpha},v_{\beta})$ denote the position of the equally
spaced grid points. We then split equation (\ref{eq:L}) into
\begin{equation}
\partial_{t}L_{i}^{f}=-v\partial_{z}L_{i}^{f}\label{eq:L-subflow1}
\end{equation}

and 
\begin{equation}
\partial_{t}L_{i}^{f}=-\frac{1}{M_{e}}\sum_{k}(e_{ik}+e_{ik}^{A})\partial_{v}L_{k}^{f}.\label{eq:subflow2}
\end{equation}

Since in equation (\ref{eq:L-subflow1}) the speed of the advection
does not depend on the advection variables, we can easily solve this
exactly in Fourier space (up to the truncation error of the Fourier
series). This yields
\[
L_{i;\cdot\beta}^{f}(\Delta t)=\text{FFT}^{-1}\left(\exp(-iv_{\beta}k_{z}\Delta t)\bullet\text{FFT}\bigl(L_{i;\cdot\beta}^{f}(0)\bigr)\right),
\]
where $k_{z}$ is the vector that contains all the frequencies in
the $z$-direction and $\bullet$ denotes the component-wise product
of two vectors. 

For equation (\ref{eq:subflow2}) the situation is more complicated.
This is primarily due to the fact that the different $L_{i}^{f}$
couple to each other. However, since the matrix $E_{ik}(z)=e_{ik}(z)+e_{ik}^{A}(z)$
is symmetric (for each fixed $z$) we can diagonalize it, i.e.
\[
\text{diag}(\lambda)=T^{T}ET.
\]

Then we perform a coordinate transformation into the Eigenbasis
\[
M_{i}^{f}=\sum_{n}T_{in}^{T}L_{n}^{f}.
\]
which we use in equation (\ref{eq:subflow2}) to obtain
\begin{align*}
\partial_{t}M_{i}^{f} & =-\frac{1}{M_{e}}\lambda_{i}\partial_{v}M_{i}^{f}.
\end{align*}
This can now be treated exactly (up to the truncation error) in Fourier
space

\[
M_{i;\alpha\cdot}^{f}(\Delta t)=\text{FFT}^{-1}\left(\exp(-i\lambda_{i;\alpha}k_{v}\Delta t/M_{e})\bullet\text{FFT}\bigl(M_{i;\alpha\cdot}^{f}(0)\bigr)\right),
\]
where $k_{v}$ are the frequencies in the $v$-direction and we note
that the eigenvalues depend on $z$ and thus $\alpha$ once the discretization
has been performed. We now have to simply undo the coordinate transformation
to complete the algorithm and obtain $L_{i;\alpha\beta}^{f}(\Delta t)$.

We have chosen a Fourier spectral method here. However, it should
be noted that any semi-Lagrangian method could be used in its place
(as has commonly been done for kinetic problems; see, e.g., \citep{Sonnendruecker1999,Grandgirard2006,Latu2014,Einkemmer2019b}).

\emph{Lax--Wendroff discretization: }As an alternative we can also
employ a more traditional Lax--Wendroff scheme. For equation (\ref{eq:L-subflow1})
this is immediate. For equation (\ref{eq:subflow2}) we again change
to the Eigenbasis and then apply the Lax--Wendroff method, i.e.
\[
M_{i;\alpha\beta}^{f;1}=M_{i}^{f;0}-\frac{\Delta t}{2\Delta x}\frac{\lambda_{i\alpha}}{M_{e}}\left(M_{i;\alpha,\beta+1}^{f;0}-M_{i;\alpha,\beta-1}^{f;0}\right)+\frac{\Delta t^{2}}{2\Delta x^{2}}\frac{\lambda_{i;\alpha}^{2}}{M_{e}^{2}}\left(M_{i;\alpha,\beta+1}^{f;0}-2M_{i;\alpha\beta}^{f;0}+M_{i;\alpha,\beta-1}^{f;0}\right),
\]
where $M_{i;\alpha\beta}^{f;1}\approx M_{i}^{f}(\Delta t,z_{\alpha},v_{\beta})$
and $M_{i;\alpha\beta}^{f;0}\approx M_{i}^{f}(0,z_{\alpha},v_{\beta})$. 

\subsection{Adaptive step size controller\label{subsec:adaptive-step-size}}

In the numerical results presented in this paper we will use an adaptive
step size controller. This has the advantage that the user is freed
from choosing an appropriate time step size. For the standard approach
we refer to \citep[Chap. II.4]{hairer1993} and \citep{Gustafsson1988},
but note that is has been realized that, in particular, for implicit
and exponential methods more efficient controllers can be constructed
(see, e.g., \citep{Gustafsson1997,Caliari2016,Einkemmer2018a,Deka2022,Deka2022a}).

Since we use explicit and relatively low order methods here, we follow
the classic idea of computing the solution once with time step $\Delta t$
and once with time step $\Delta t/2$ and then estimating the error
using Richardson extrapolation. An important question in this context
is the appropriate metric for the error. We take the viewpoint that
in many physical problems one is primarily interested in the evolution
of macroscopic quantities and not necessarily in all of the details
of the particle-density function. Thus, we consider for the step size
controller the error in the electric and magnetic energy. This has
the added benefit that those quantities can be computed from the low-rank
representation at very low additional cost.

More specifically, the numerical scheme with time step size $\Delta t$
yields a electric energy at time $t^{n+1}$ denoted by $\text{ee}_{\Delta t}^{n+1}$
and a magnetic energy denoted by $\text{me}_{\Delta t}^{n+1}$, see
equations (\ref{eq:ee}) and (\ref{eq:me}) for a definition. Similar
the numerical scheme applied twice with time step size $\Delta t/2$
yields electric energy at time $t^{n+1}$ denoted by $\text{ee}_{\Delta t/2}^{n+1}$
and magnetic energy denoted by $\text{me}_{\Delta t/2}^{n+1}$. From
this we can obtain an estimate using Richardson extrapolation
\[
\text{\text{ee}}_{\text{RE}}^{n+1}=\frac{2^{p+1}\text{ee}_{\Delta t/2}^{n+1}-\text{ee}_{\Delta t}^{n+1}}{2^{p+1}-1},\qquad\text{\text{me}}_{\text{RE}}^{n+1}=\frac{2^{p+1}\text{me}_{\Delta t/2}^{n+1}-\text{me}_{\Delta t}^{n+1}}{2^{p+1}-1},
\]
where the local truncation error of the scheme is assumed to be $\mathcal{O}(\Delta t^{p+1})$.
We have $p=2$ for the Strang based projector splitting and $p=1$
for all other methods considered here. The Richardson extrapolated
value has local truncation error of size $\mathcal{O}(\Delta t^{p+2})$
and can thus be used to estimate the error of $\text{ee}_{\Delta t/2}^{n+1}$
and $\text{me}_{\Delta t/2}^{n+1}$. We consider

\begin{align*}
\left(\text{err}^{n+1}\right)^{2} & =(\text{ee}_{\Delta t/2}^{n+1}-\text{ee}_{RE}^{n+1})^{2}+(\text{me}_{\Delta t/2}^{n+1}-\text{me}_{RE}^{n+1})^{2}\\
 & =\frac{(\text{ee}_{\Delta t}^{n+1}-\text{ee}_{\Delta t/2}^{n+1})^{2}+(\text{me}_{\Delta t}^{n+1}-\text{me}_{\Delta t/2}^{n+1})^{2}}{(2^{p+1}-1)^{2}}.
\end{align*}
This is the absolute error. For the step size controller we use the
relative error per unit time step. That is,
\[
\text{rel\_err\_unit}^{n+1}=\frac{\text{err}^{n+1}}{\Delta t\sqrt{(\text{ee}_{\Delta t/2}^{n+1})^{2}+(\text{me}_{\Delta t/2}^{n+1})^{2}}}.
\]
The new step size is then determined according to the classic controller
\begin{equation}
\Delta t^{\text{new}}=\Delta t\left(s\frac{\text{tol}}{\text{\ensuremath{\text{rel\_err\_unit}^{n+1}}}}\right)^{1/(p+1)},\label{eq:new-step-size}
\end{equation}
where $s$ is a safety factor ($s=0.7$ is chosen in the numerical
experiments). If $\text{rel\_err\_unit}^{n+1}>\text{tol}$ the step
is rejected and the time step is repeated either with the new step
size chosen according to equation (\ref{eq:new-step-size}) or half
the previous time step, whichever of those two is smaller.

\section{Properties of the dynamical low-rank algorithm\label{sec:properties-dlr-algo}}

In order to show that the dynamical low-rank algorithm, for a given
problem and initial value, works well, we need two ingredients. First,
we need to show that the solution can be represented well by a low-rank
approximation. That is, that $(f,\phi,A)$ can be approximated accurately
by equations (\ref{eq:lr-f}) and (\ref{eq:lr-phiA}) with a small
rank $r$. Second, we need to show that the dynamical low-rank algorithm
approximates this solution accurately. In this section we will show
that this is the case if the dynamics is not too far from linear theory. 

We start by linearizing equation (\ref{eq:drift-kinetic}) at $(f,\phi,A)=(f_{eq},0,0)$
and obtain
\begin{equation}
\partial_{t}f+v\partial_{z}f+\frac{1}{M_{e}}\left(\partial_{z}\phi+\partial_{t}A\right)\partial_{v}f_{eq}=0,\label{eq:linearized-driftkinetic}
\end{equation}
where we again use $f$ to denote the perturbation from $f_{eq}$.
From this, the dispersion relation of the problem can be obtained
(the detailed calculations can be found in \ref{app:dispersion}).
The dispersion relation tells us that if initially a single spatial
mode with wavenumber $k=(k_{x},k_{y},k_{\parallel})$ is excited,
i.e.~we consider the initial value
\[
f(0,x,v)=(1+\alpha\cos(k_{x}x)\cos(k_{y}y)\cos(k_{\parallel}z))f_{eq}(v),
\]
where $f_{eq}$ is a Maxwellian and $\alpha$ is the strength of the
perturbation, the solution at later time can be written as
\begin{equation}
f(t,x,v)=\text{Re}(1+\alpha\exp(ik\cdot x-\omega t))f_{eq}(v),\;\;\phi(t,x)\propto\text{Re}\exp(ik\cdot x-\omega t),\;\;A(t,x)\propto\text{Re}\exp(ik\cdot x-\omega t),\label{eq:lin-sol}
\end{equation}

where $\omega$ is determined by
\begin{equation}
1-\frac{2[1+\overline{\omega}Z(\overline{\omega})]}{\left(k_{\perp}\rho_{i}/L\right)^{2}}(\beta/M_{e}\overline{\omega}^{2}-1)=0,\qquad\qquad\overline{\omega}=\sqrt{M_{e}}\frac{\omega}{k_{\parallel}}.\label{eq:dispersion-relation}
\end{equation}

The important observation here is that this is a rank $1$ solution.
Thus, the solution has the correct structure to be approximated efficiently
by equations (\ref{eq:lr-f}) and (\ref{eq:lr-phiA}).

However, this on its own is not sufficient. We also need that the
dynamical low-rank algorithm, i.e.~the equations of motions (\ref{eq:K}),
(\ref{eq:S}), and (\ref{eq:L}), are able to capture this low-rank
solution. Ideally, we want an algorithm that if the solution of our
problem has a certain rank, say $s$, and we run the simulation with
a rank $r\geq s$ then we obtain the exact result (up to time and
space discretization errors).

For the dynamical low-rank approach this is indeed the case. In fact,
we can write the dynamical low-rank approximation applied to equation
(\ref{eq:linearized-driftkinetic}) as
\begin{equation}
\partial_{t}f=P(f)\text{LIN}(f)=P(f)\left(-v\partial_{z}f-\frac{1}{M_{e}}\left(\partial_{z}\phi+\partial_{t}A\right)\partial_{v}f_{eq}\right),\label{eq:lowrank-linear}
\end{equation}
where $P(f)$ is the orthogonal projector to the tangent space of
functions with rank $r$ (see, e.g., \citep{Koch2007}). We also know
that equation (\ref{eq:lin-sol}) is a solution to $\partial_{t}f=\text{LIN}(f)$.
However, since $f$ is rank $1$, $\text{LIN}(f)$ lies in the corresponding
tangent space and the projection operator is just the identity. That
is, we have $P(f)\text{LIN}(f)=\text{LIN}(f)$ and the low-rank solution
given by equation (\ref{eq:lin-sol}) is therefore also a solution
of equation (\ref{eq:lowrank-linear}). Thus, within linear theory
the low-rank algorithm is exact as the equations of motions (\ref{eq:K}),
(\ref{eq:S}), and (\ref{eq:L}) are precisely the same as equation
(\ref{eq:lowrank-linear}), only expressed in the low-rank factors
$X$, $S$, and $V$ instead of $f$. 

We should duly note that this argument only applies to the dynamical
low-rank algorithm as described in sections (\ref{subsec:evolution})-(\ref{subsec:dtA}).
In general, we have to apply a time and space discretization and those
introduce numerical errors. However, this is also true for a direct
discretization of the full problem. The only specific issue for dynamical
low-rank approximations is that the robust integrators we use (see
section \ref{subsec:dlr-algorithm}) also introduces an additional
time discretization error. This error tends to zero as $\Delta t$
tends to zero (as is the case for any reasonable time discretization).

\section{Alfvén waves \label{sec:numerical-results}}

We will now use the proposed dynamical low-rank algorithm to simulate
the propagation of kinetic shear Alfvén waves. We will also compare
the results both to analytic theory (in cases where it applies) and
a code that directly discretizes the full four-dimensional problem,
i.e.~equation (\ref{eq:drift-kinetic}). As initial value we consider
a Maxwellian equilibrium of the electrons that is perturbed in all
spatial directions, i.e.

\begin{equation}
f(0,x,y,z,v)=\left(1+\alpha\cos(k_{x}x)\cos(k_{y}y)\cos(k_{\parallel}z)\right)\frac{\exp(-M_{e}v^{2})}{\sqrt{\pi/M_{e}}}\label{eq:iv}
\end{equation}
on the domain $(x,y,z,v)\in[0,2\pi/k_{x})\times[0,2\pi/k_{y})\times[0,2\pi/k_{\parallel})\times[-6/\sqrt{M_{e}},6/\sqrt{M_{e}})$.
We use the physical mass ratio $M_{e}=1/1830$ and choose $\beta/M_{e}=1.8$.
The wavevectors $k_{x}$, $k_{y}$, and $k_{z}$ of the perturbation
in the $x$, $y$, and $z$ directions determine according to the
dispersion relation (see \ref{app:dispersion}), assuming the linear
theory holds true, the decay rate and frequency of the wave. We consider
here $k_{x}=k_{y}=k_{\perp}/\sqrt{2}$ with $k_{\perp}\rho_{i}=0.2$,
$k_{\parallel}=2\pi$, and $\alpha=10^{-5}$. Note that for the dynamics
only the value of the dimensionless parameter $k_{\perp}\rho_{i}$
is important and not how $k_{\perp}$ and $\rho_{i}$ is chosen individually.
For this configuration the dispersion relation gives a decay rate
of $\gamma\approx2.4016$ and a angular frequency of $\omega\approx201.034$. 

The expected physical behavior in the linear regime (see, e.g., \citep{Dannert2004,Lu2021})
is that the wave trades energy between the electric and magnetic field,
while the amplitude of the field quantities decay in time. The energy
lost from the fields is transferred to the kinetic energy of the plasma.
In the following we will consider the electric energy
\begin{equation}
\text{ee}=\frac{1}{2C_{P}}\int\Vert(\partial_{xx}+\partial_{yy})\phi\Vert^{2}\,d(x,y,z)\label{eq:ee}
\end{equation}

magnetic energy
\begin{equation}
\text{me}=\frac{1}{2C_{A}}\int\Vert(\partial_{xx}+\partial_{yy})A\Vert^{2}\,d(x,y,z)\label{eq:me}
\end{equation}
and the kinetic energy
\[
\text{ke}=\frac{M_{e}}{2}\int v^{2}f\,d(x,y,z,v).
\]
We know that the total energy, i.e. $\text{ee}+\text{me}+\text{ke}$,
is an invariant of equation (\ref{eq:drift-kinetic}). In addition,
to the total energy, we will also consider the total mass
\[
\int f\,d(x,y,z,v)
\]
and the total momentum
\[
\int vf\,d(x,y,z,v),
\]
which are also invariants of the dynamics, in order to judge the quality
of the numerical solution obtained. To compute the relative error
for mass and energy we normalize with respect to the mass and energy
of the initial value. The momentum is initially zero and we thus normalize
with respect to $\int\vert v\vert f(0,x,y,z,v)\,d(x,y,z,v)$. 

\begin{figure}[H]
\begin{centering}
\includegraphics[width=16cm]{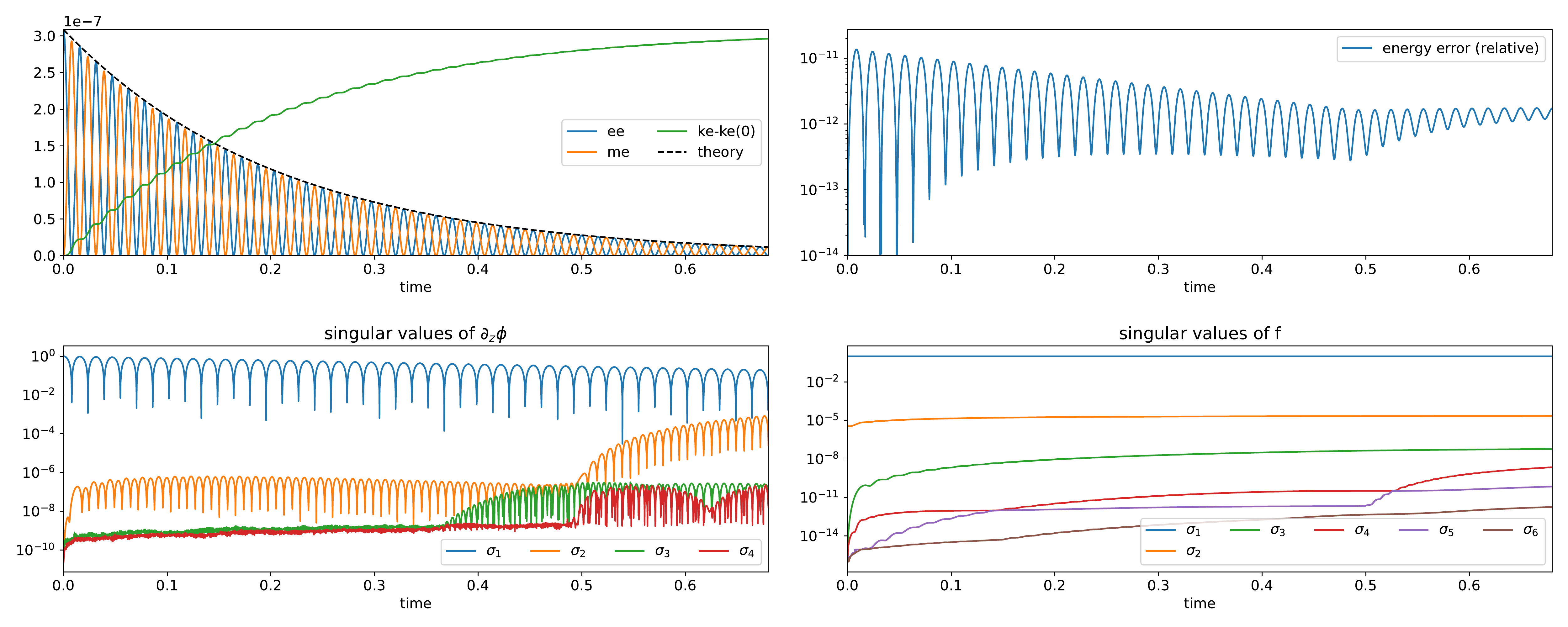}
\par\end{centering}
\caption{Numerical simulations of equation (\ref{eq:drift-kinetic}) with initial
value given by equation (\ref{eq:iv}) with $\beta/M_{e}=1.8$ and
$\alpha=10^{-5}$ are shown. An FFT based direct (i.e.~full rank)
solver with time step size $\Delta t=10^{-4}$ and $25$, $25$, $25$,
and $512$ grid points in the $x$, $y$, $z$, and $v$ direction,
respectively, is used in all plots. On the top-left the time evolution
of the electric energy, magnetic energy, and kinetic energy is shown.
The analytic decay rate $\gamma\approx2.4016$ is illustrated using
a dashed black line. On the top right the violation of energy conservation
by the algorithm is shown. On the bottom we show the magnitude of
the dominant singular values for $\partial_{z}\phi$ (left) and the
particle-density function $f$ (right). All singular values are normalized
with respect to the largest singular value at time $t=0$. \label{fig:full-alpha1e-5}}
\end{figure}

We first present numerical results using a solver that directly discretizes
equation (\ref{eq:drift-kinetic}), henceforth called the full rank
simulation. The results in Figure \ref{fig:full-alpha1e-5} show excellent
agreement of the numerical solution compared to the decay rate and
frequency predicted from linear theory, as we expect for the relatively
small perturbation considered. We also use this simulation to investigate
the low-rank structure of the simulation. To do that we perform a
singular value decomposition for both $f$ and $E$ at each time step
and plot the magnitude of the singular values. As we can see, from
Figure \ref{fig:full-alpha1e-5}, the dynamics is dominated by a single
low-rank mode. Due to nonlinear effects some additional low-rank modes
are excited. However, those are orders of magnitude smaller in magnitude
than the dominant low-rank mode. Thus, the structure of the solution
is inherently low-rank and can be well represented by such an approximation
(as we would expect based on the discussion in section \ref{sec:properties-dlr-algo}).

\begin{figure}[H]
\begin{centering}
\includegraphics[width=16cm]{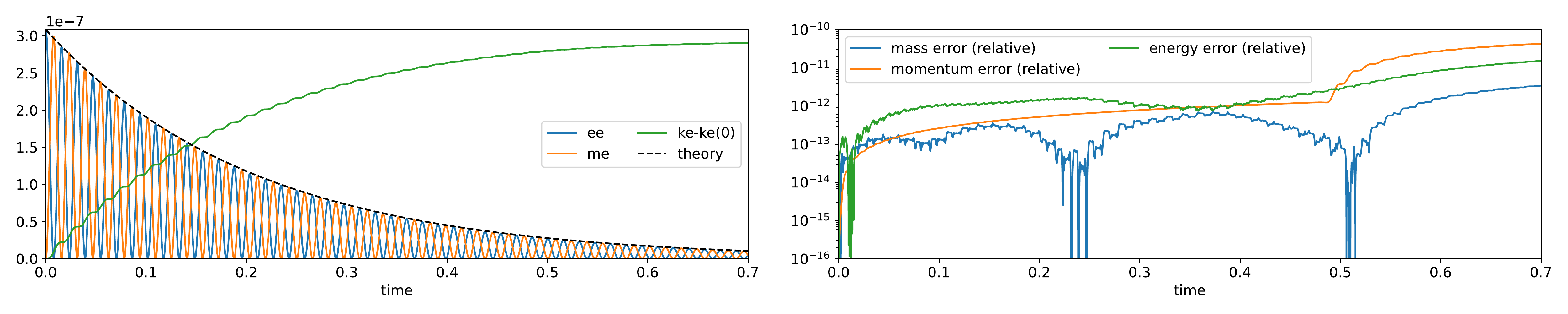}
\par\end{centering}
\begin{centering}
\includegraphics[width=16cm]{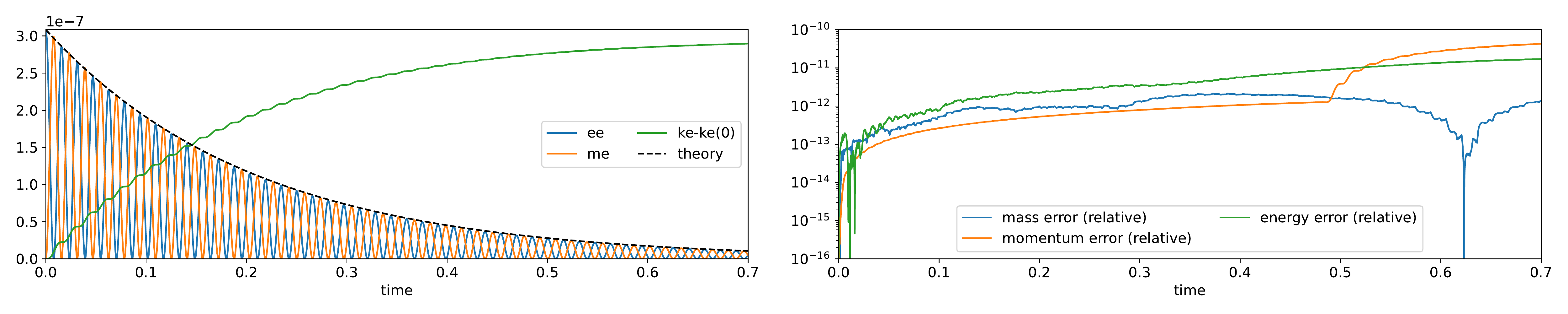}
\par\end{centering}
\caption{Numerical simulations of equation (\ref{eq:drift-kinetic}) with initial
value given by equation (\ref{eq:iv}) with $\beta/M_{e}=1.8$ and
$\alpha=10^{-5}$ are shown. The dynamical low-rank Lie projector
splitting with $r=2$ (top) and $r=5$ (bottom) is employed with an
adaptive time step controller (the tolerance is set to $10^{-1}$)
and 32, 32, 32, and 512 grid points in the $x$, $y$, $z$, and $v$
directions, respectively. On the left, the time evolution of the electric
energy, magnetic energy, and kinetic energy is shown. The analytic
decay rate $\gamma\approx2.4016$ is illustrated using a dashed black
line. The violation of mass, momentum, and energy conservation by
the dynamical low-rank algorithm is shown on the right. \label{fig:lr-r5-alpha1e-5}}
\end{figure}

We now consider the dynamical low-rank algorithm proposed in this
paper. Details of the implementation and its computational efficiency
will be discussed in section \ref{sec:comparison}. We consider the
same configuration as outlined above and simulations with fixed ranks
$r=2$ and $r=5$ are conducted. The results are shown in Figure \ref{fig:lr-r5-alpha1e-5}.
We observe excellent agreement with the linear theory and with the
full rank simulation. Moreover, the error in mass is close to machine
precision and the error in momentum is below $10^{-10}$. The error
in energy, approximately $10^{-11}$, has a similar magnitude than
for the direct solver considered in Figure \ref{fig:full-alpha1e-5}.
This indicates that the space and time discretization has a more significant
effect on the error in energy than the low-rank approximation. Thus
we conclude that the dynamical low-rank algorithm, even at extremely
small ranks such as $r=2$, can obtain comparable results to the full
rank simulation at drastically reduced computational and memory cost.

\begin{figure}[H]
\begin{centering}
\includegraphics[width=16cm]{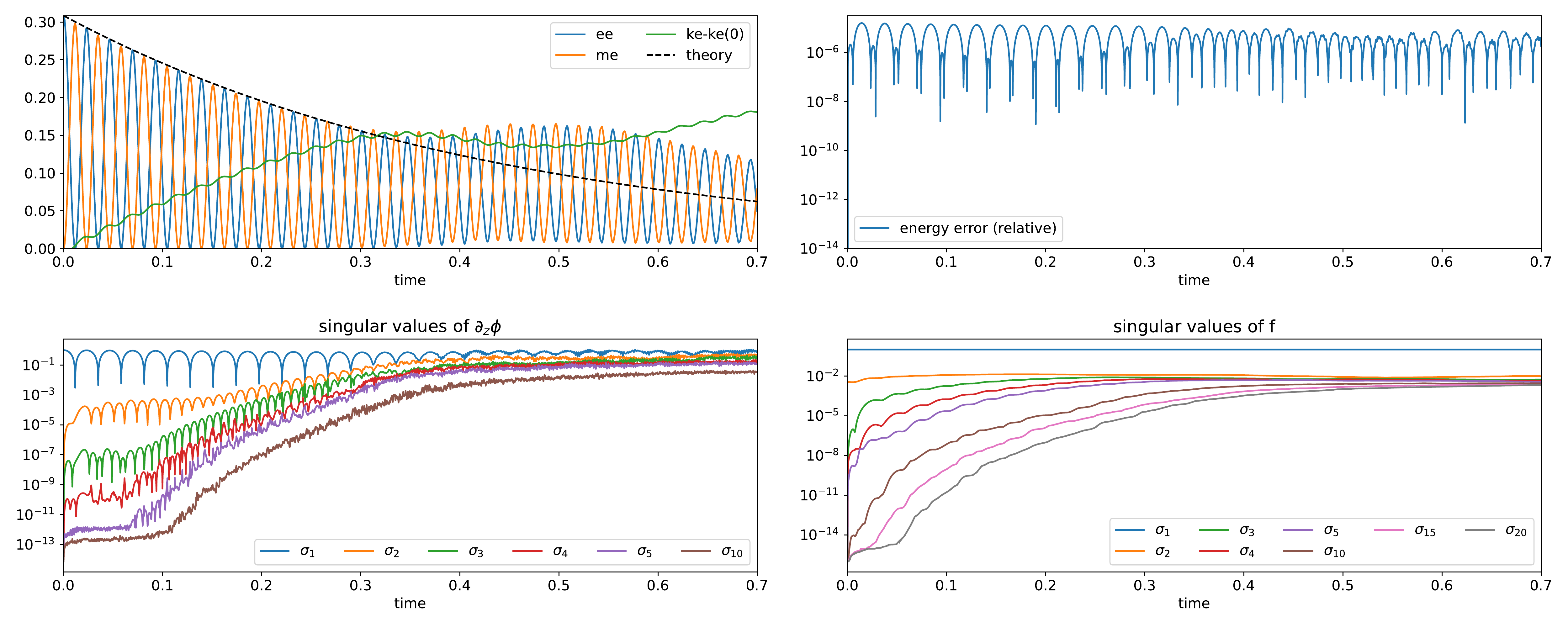}
\par\end{centering}
\caption{Numerical simulations of equation (\ref{eq:drift-kinetic}) with initial
value given by equation (\ref{eq:iv}) with $\beta/M_{e}=4$ and $\alpha=10^{-2}$
are shown. An FFT based direct (i.e.~full rank) solver with time
step size $\Delta t=10^{-4}$ and $25$, $25$, $25$, and $512$
grid points in the $x$, $y$, $z$, and $v$ direction, respectively,
is used in all plots. On the top-left the time evolution of the electric
energy, magnetic energy, and kinetic energy is shown. The analytic
decay rate $\gamma\approx1.1413$ is illustrated using a dashed black
line. On the top right the violation of energy conservation by the
algorithm is shown. On the bottom we show the magnitude of the dominant
singular values for $\partial_{z}\phi$ (left) and the particle-density
function $f$ (right). All singular values are normalized with respect
to the largest singular value at time $t=0$. \label{fig:full-alpha1e-2}}
\end{figure}

Second, we consider $\beta/M_{e}=4$ and a significantly larger perturbation
$\alpha=10^{-2}$. From the full rank simulation in Figure \ref{fig:full-alpha1e-2}
we observe an initial decay of the energy stored in the field quantities
that matches the linear theory well. However, at approximately $t=0.3$
nonlinear effects take over and the decay of the energy stored in
the field quantities is stopped. These results match well with what
has been reported in the literature (see, e.g. \citep{Lu2021}; we
note that for the present configuration and the used normalization
the Alfvén time is $T_{\text{SAW}}=2\pi/(v_{A}k_{\parallel})=\sqrt{\beta}\approx0.047$
and thus the simulation time covers approximately $15T_{\text{SAW}}$). 

\begin{figure}[H]
\begin{centering}
\includegraphics[width=16cm]{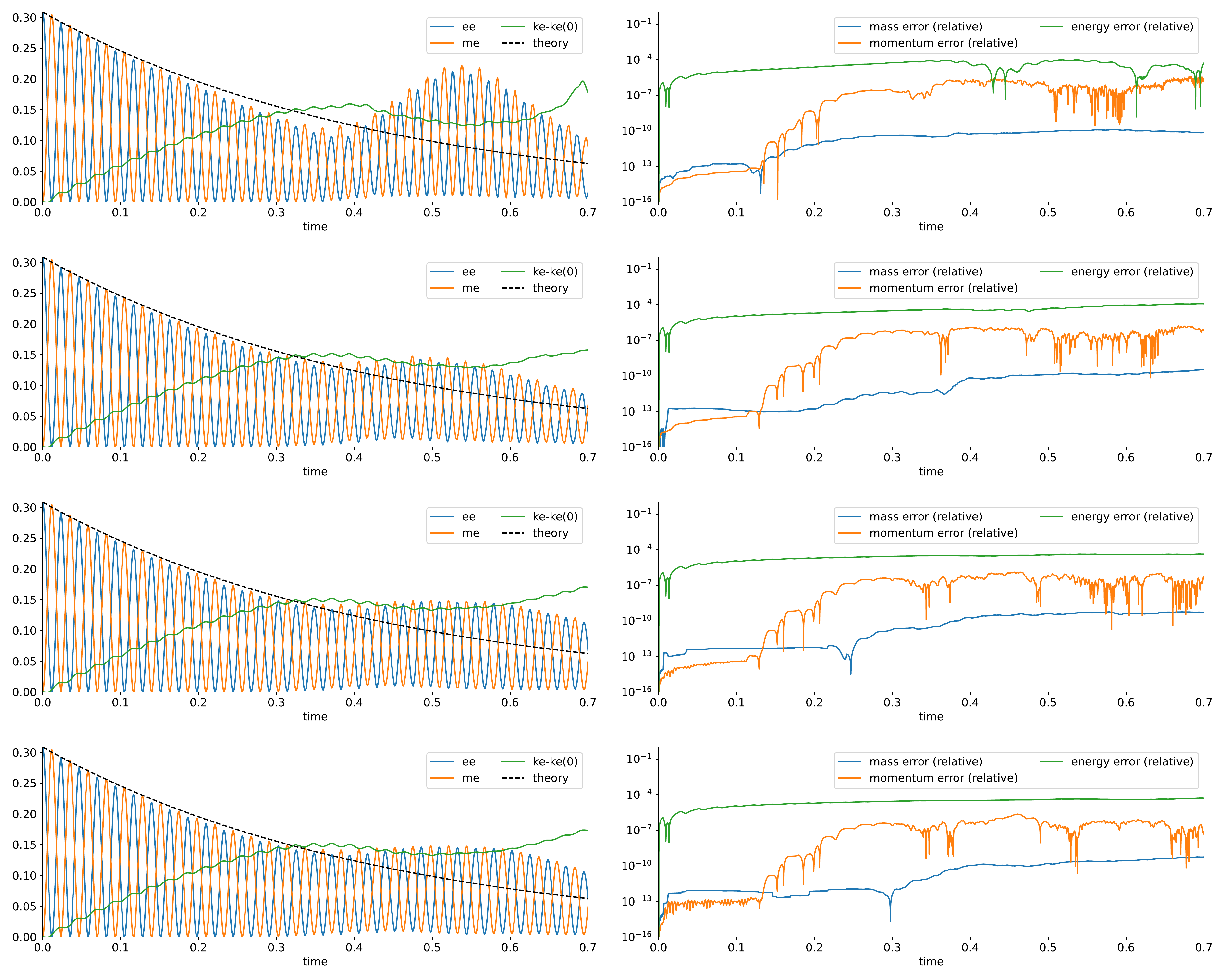}
\par\end{centering}
\caption{Numerical simulations of equation (\ref{eq:drift-kinetic}) with initial
value given by equation (\ref{eq:iv}) with $\beta/M_{e}=4$ and $\alpha=10^{-2}$
are shown. The Strang projector splitting dynamical low-rank approach
proposed in this paper with (from top to bottom) $r=3$, $r=5$, $r=8$,
and $r=10$ is employed with an adaptive time step controller (the
tolerance is set to $10^{-1}$) and 32, 32, 32, and 512 grid points
in the $x$, $y$, $z$, and $v$ directions, respectively. On the
left the time evolution of the electric energy, magnetic energy, and
kinetic energy is shown. The analytic decay rate $\gamma\approx1.1413$
is illustrated using a dashed black line. The violation of mass, momentum,
and energy conservation by the dynamical low-rank algorithm is shown
on the right.\label{fig:lr-alpha1e-2}}
\end{figure}

We now turn our attention to the low-rank structure of the solution.
Initially, the solution is dominated by a small number of low-rank
modes, as we would expect in the regime where the linear theory holds.
However, as time increases and nonlinear effects become more important,
in addition to the dominant mode there is a multitude of low-rank
modes with singular values with size approximately on the order of
$10^{-2}$. Thus, it is unclear how large the rank needs to be chosen
in order to obtain a good approximation to the solution. In Figure
\ref{fig:lr-alpha1e-2} we did run the dynamical low-rank algorithm
with rank $r=3$, $r=5$, $r=8$, and $r=10$. For rank $r=3$ the
results are at most qualitatively correct (we observe a wave with
roughly the right frequency, but the amplitude is significantly larger
than for the direct simulation). However, starting with rank $r=5$
we obtain results that match the result from the full rank simulation
very well. There is only a slight improvement going to rank $r=8$.
The simulation for $r=8$ and $r=10$ are almost indistinguishable.
We also note that for all configurations the error in energy is of
a similar magnitude as in the full rank simulation, suggesting that
this error is dominated by the space and time discretization.

\section{Comparison of numerical methods\label{sec:comparison}}

In order to take the (continuous) dynamical low-rank approximation
described in sections \ref{subsec:evolution}-\ref{subsec:dtA} and
put it on a computer we have choose a robust integrator (see section
\ref{subsec:dlr-algorithm}) and perform a space and time discretization
(see section \ref{subsec:discretization}). In this section we will
consider how those choices effect computational performance. Our implementation
of the dynamical low-rank algorithm is based on the open-source \texttt{C++}
framework Ensign~\citep{Cassini2021}\footnote{Ensign is available at \url{https://github.com/leinkemmer/Ensign}
under the MIT license.} and the code used in this paper and the corresponding unit tests
can be found in the \texttt{examples }subfolder of that software package.

\begin{figure}[H]
\begin{centering}
\includegraphics[width=16cm]{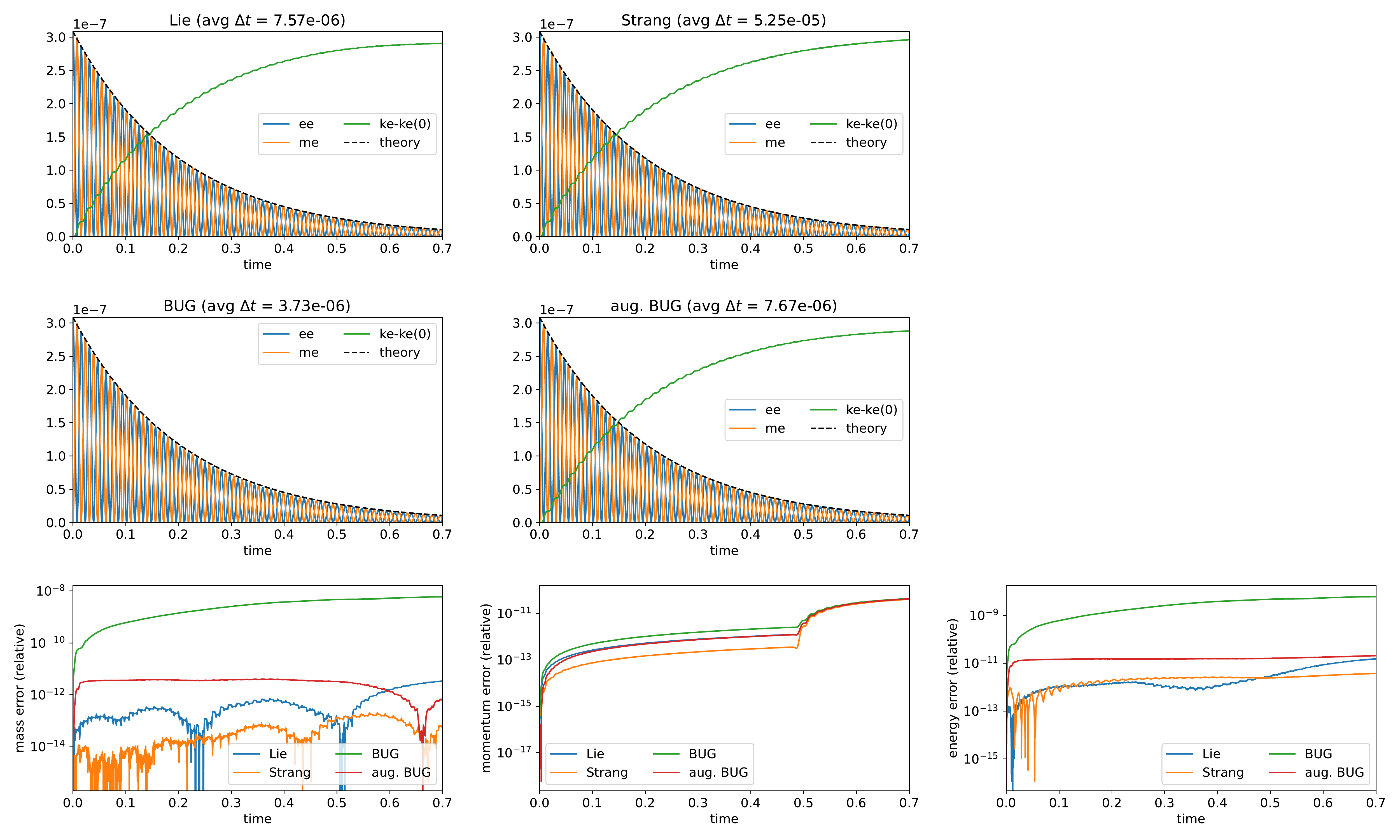}
\par\end{centering}
\caption{Time evolution of the electric, magnetic, and kinetic energy is shown
for the Lie projector splitting (top-left), the Strang projector splitting
(top-right), the BUG integrator (middle-left), and the augmented BUG
integrator (middle-right). In all simulations the adapive time stepping
algorithm described in section \ref{subsec:adaptive-step-size} with
a tolerance of $10^{-1}$ is used. On the bottom the error in mass
(left), momentum (middle), and energy (right) committed by these methods
is shown. The rank in all simulations is set to $r=2$. \label{fig:robust-integrators}}
\end{figure}

First, let us consider the choice of the robust integrator. We have
three first order methods, the Lie projector splitting integrator
(Algorithm \ref{alg:projector-lie}), the BUG or unconventional integrator
(Algorithm \ref{alg:bug}), and the augmented BUG integrator (Algorithm
\ref{alg:augmented-bug}). We also consider the second order Strang
projector splitting integrator (Algorithm \ref{alg:projector-strang}).
In Figure \ref{fig:robust-integrators} we compare those four integrators
for the configuration with $\beta/M_{e}=1.8$ and $\alpha=10^{-5}$
considered in the previous section. We employ the adaptive time stepping
algorithm described in section \ref{subsec:adaptive-step-size}. All
integrators yield an electric and magnetic energy that matches the
theoretical value and the full rank simulation well. Note, however,
that the step size at which this is achieved is significantly different.
For the second order Strang splitting we can take an (average) step
size which is approximately a factor of $7$ larger. Thus, more than
making up for the fact that Strang splitting is approximately two
times as costly as Lie splitting. As we would expect, conservation
of the invariants, in particular, energy is also better for Strang
splitting as compared to the other robust integrators. The Lie splitting
scheme is approximately able to take the same time step size as the
augmented BUG integrator, while being significantly less expensive
(see Table \ref{tab:timing}, which we discuss in more detail later
in this section).

The outlier here is the BUG integrator. To obtain accurate results
for the electric and magnetic energy we require more than twice as
many time steps as for the first order projector splitting or the
augmented BUG integrator. It also performs much worse with respect
to mass and energy conservation (the error is approximately three
orders of magnitude larger than for the projector splitting or the
augmented BUG integrator). 

\begin{figure}[H]
\begin{centering}
\includegraphics[width=16cm]{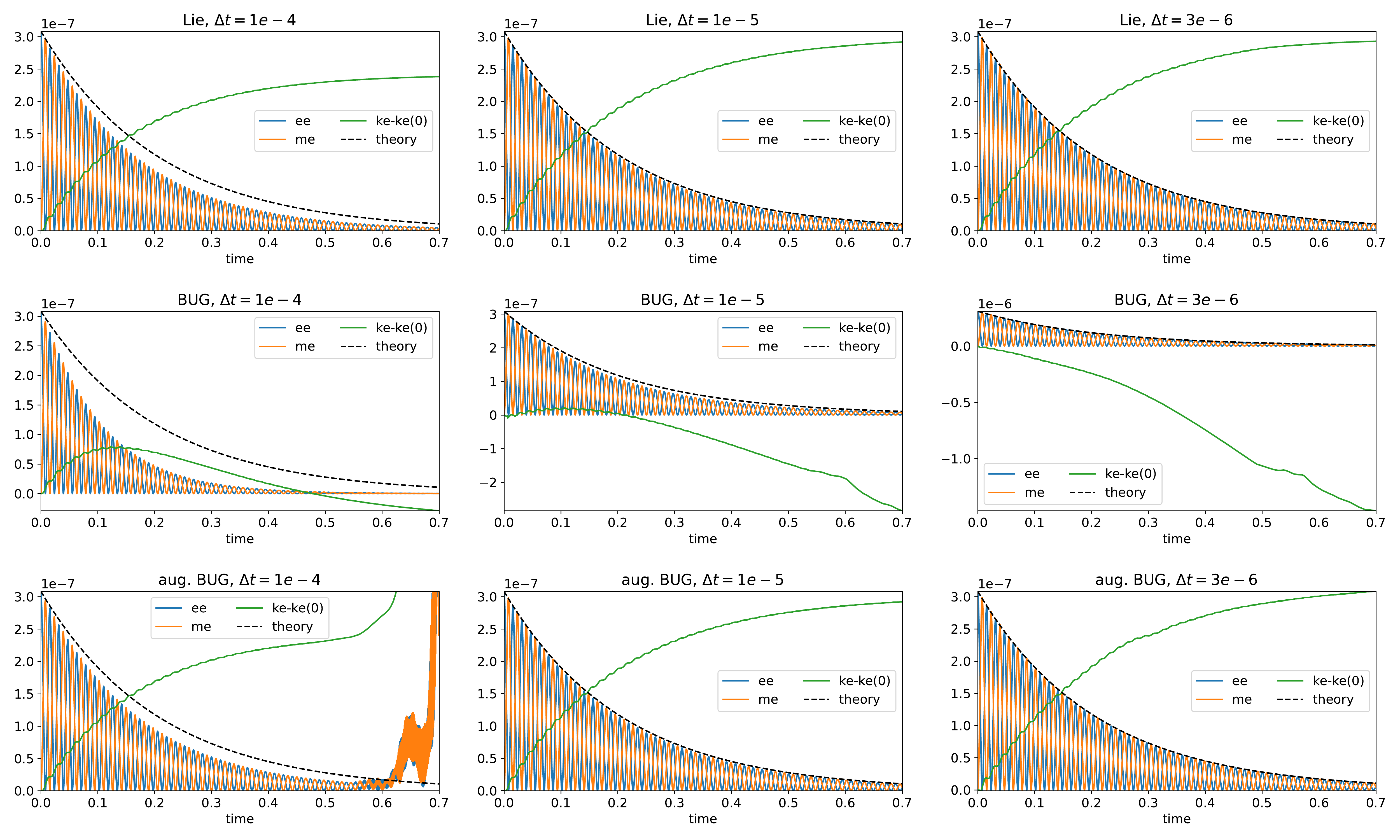}
\par\end{centering}
\caption{Time evolution of the electric, magnetic, and kinetic energy for the
three first order integrators for three different fixed time step
sizes is shown. Each column corresponds to a different time step size
and each row corresponds to a different method. The rank in all simulations
is set to $r=5$. \label{fig:fixed-deltat}}
\end{figure}

To investigate this further we did run the three first order schemes
with different fixed time step sizes. The result is shown in Figure
\ref{fig:fixed-deltat} (the time step size decreases as we go from
left to right). The most striking is that the BUG integrator actually
decreases the kinetic energy instead of increasing it to balance out
the reduction in electric and magnetic energy. This is a consequence
of the bad energy conservation properties of that integrator that
we observed earlier. Neither the projector splitting nor the augmented
BUG integrator have this issue. As we increase the time step size
both the projector splitting and the augmented BUG integrator become
less accurate, as we would expect. However, for the largest time step
size considered in Figure \ref{fig:fixed-deltat} the augmented BUG
integrator becomes unstable, whereas the Lie projector splitting integrator
still gives qualitatively correct. 

Let us now consider the computational efficiency of the different
low-rank algorithms. Both the run time for different configurations
(i.e.~different rank $r$ and different spatial discretizations)
as well as the memory consumption is shown in Table \ref{tab:timing}.
We compare our C++ low-rank implementation to a direct (i.e.~full
rank) solver. The direct solver is implemented in Julia and thus this
should not be entirely considered an apple to apple comparison. However,
care has been taken in the Julia code to make it efficient by vectorizing
the code. Since the direct solver is relatively easy to implement,
primarily consisting of FFTs and vector operations, the performance
of the code should not be too distant from a reasonable (but perhaps
not highly optimized) C++ implementation. The low-rank code is written
in C++ using the Ensign framework. Also in this case some care has
been taken to obtain reasonable performance. In both cases we run
the code sequentially on a dual socket Intel Gold 6130 system with
192 GB of RAM. The speedup obtained obviously depends on the rank
of the simulation. For small ranks, e.g., $r=5$ (as was a reasonable
choice for all simulations considered in this paper), the speedup
of the dynamical low-rank algorithm is dramatic. Even for relatively
coarse grids (e.g.~$32^{3}\times512$ or $64^{3}\times512$) the
observed speedup is approximately three orders of magnitude. A similar
reduction in memory is also observed. The latter is important as for
fine grids (e.g.~$256^{3}\times1024$) we can not even run the full
rank simulation on a system with 192 GB of memory. However, even for
$r=15$ the dynamical low-rank algorithm only consumes approximately
$60$ MB of memory. Thus, the dynamical low-rank algorithms enables
us to easily run simulations on a workstation or even laptop computer
that would require supercomputer hardware and distributed memory parallelism
otherwise.

Table \ref{tab:timing} also contains a comparison of the different
robust dynamical low-rank integrators. The Lie integrator is the cheapest
overall, with the the BUG integrator following close behind (the primary
additional cost in the BUG integrator is the recomputation of the
coefficients in step 8 of Algorithm \ref{alg:bug}. The augmented
BUG integrator is significantly more expensive. In fact, it is approximately
equal in computational cost to the second order Strang projector splitting
integrator. This is primarily due to the coefficients for the S step
that need to be computed using rank $2r$. Overall, the Strang projector
splitting integrators, which is between two and three times as expensive
as the Lie projector splitting integrator, gives the best performance.
This is primarily due to the significantly increased time step size,
as expected for a second order scheme (note that there is currently
no second order variant of the BUG or augmented BUG integrator).
\begin{table}[H]
\begin{tabular}{rrrrrrrrr}     Resolution & rank & \shortstack{Lie\\(time/step)} & \shortstack{BUG\\(time/step)} & \shortstack{aug. BUG\\(time/step)} & \shortstack{Strang\\(time/step)} & \shortstack{speed\\up} & \shortstack{memory\\usage} & \shortstack{memory\\down} \\     \hline                 & full  &         29 s & -- & -- & -- & -- &       0.27 GB & -- \\                 &     2  &      0.013 s &      0.015 s &       0.02 s &      0.035 s &       2272 &          1 MB &        512 \\ $32^3 \times 512$ &     5  &      0.042 s &      0.056 s &      0.091 s &       0.13 s &        683 &          1 MB &        205 \\                 &    10  &        0.2 s &       0.26 s &       0.61 s &       0.63 s &        146 &          3 MB &        102 \\                 &    15  &       0.74 s &       0.91 s &        3.3 s &        2.2 s &         39 &          4 MB &         68 \\ \hline                 & full  &    2.3e+02 s & -- & -- & -- & -- &        2.1 GB & -- \\                 &     2  &      0.024 s &      0.038 s &      0.043 s &      0.075 s &       9276 &          1 MB &       2048 \\ $64^3 \times 512$ &     5  &       0.13 s &       0.16 s &       0.26 s &       0.35 s &       1764 &          3 MB &        819 \\                 &    10  &        0.8 s &       0.87 s &        1.9 s &          2 s &        283 &          5 MB &        410 \\                 &    15  &        2.9 s &          3 s &          8 s &          7 s &         78 &          8 MB &        273 \\ \hline                 & full  & -- & -- & -- & -- & -- &    2.7e+02 GB & -- \\                 &     2  &       0.45 s &        0.5 s &       0.58 s &        1.1 s & --  &          8 MB &      32768 \\ $256^3 \times 1024$ &     5  &        4.1 s &        4.2 s &        6.7 s &        9.2 s & --  &         21 MB &      13107 \\                 &    10  &         35 s &         34 s &         70 s &         76 s & --  &         42 MB &       6554 \\                 &    15  &    1.7e+02 s &    1.7e+02 s &    4.4e+02 s &    4.4e+02 s & --  &         63 MB &       4369 \\ \hline \end{tabular}

\caption{Timing result and memory consumption for the full rank simulation
and the dynamical low-rank integrator for different values of the
rank $r$ and spatial resolutions. The speedup is computed using the
Lie splitting as the full rank solution is also computed using a Lie
splitting. The values for memory consumption are computed theoretically
(for the full rank solution $2n_{x}n_{y}n_{z}n_{v}\text{sizeof(double)}$
and for the low-rank algorithm $2rn_{z}n_{v}\text{sizeof(double)}$,
where $n_{x}$, $n_{y}$, $n_{z}$, and $n_{v}$ are the number of
points in the $x$, $y$, $z$, and $v$ direction, respectively.
The memory on our compute server was not sufficient to run the full
problem using $256^{3}\times1024$ grid points. \label{tab:timing}}
\end{table}

\begin{figure}[H]
\begin{centering}
\includegraphics[width=16cm]{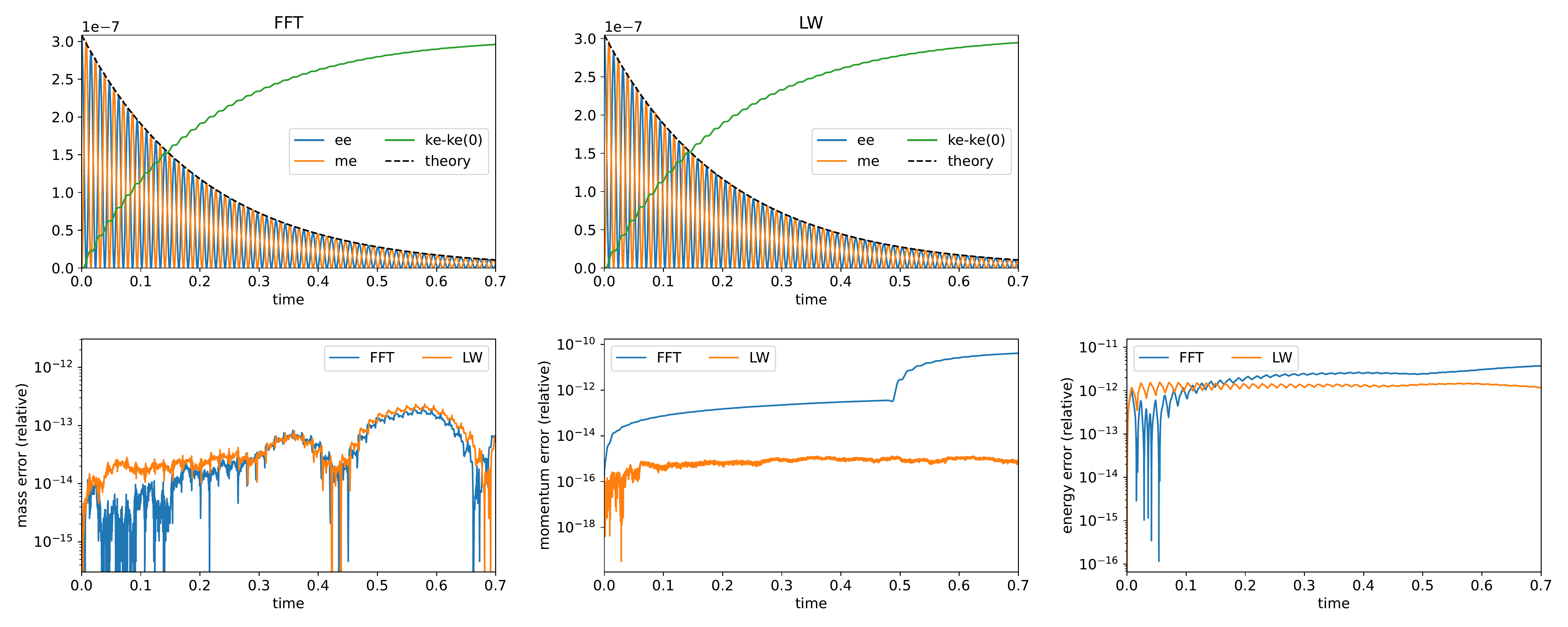}
\par\end{centering}
\caption{Time evolution of the electric, magnetic, and kinetic energy is shown
for the Strang projector splitting scheme and the FFT based spectral
discretization (top-left) and the Lax--Wendroff discretization (top-right).
On the bottom the error in mass, momentum, and energy committed by
these methods is shown. The rank in all simulations is set to $r=2$.
\label{fig:fft-vs-lw}}
\end{figure}

Let us now turn our attention to the spatial and temporal discretization.
In particular, we consider both a FFT based spectral discretization
and a discretization based on the Lax--Wendroff scheme (both methods
are described in detail in section \ref{subsec:discretization}).
The corresponding numerical results are shown in Figure \ref{fig:fft-vs-lw}.
Interestingly, the Lax--Wendroff scheme has a significantly lower
error in momentum and a sightly lower overall error in energy. Otherwise,
both methods give comparable results.

\bibliographystyle{plainnat}
\bibliography{/home/lukas/Dropbox/literature/literature}

\appendix

\section{Dispersion relation\label{app:dispersion}}

Taking Fourier transforms in time and physical space, i.e.~$u(t,x)=\sum_{\omega k}\hat{u}_{\omega k}e^{ik\cdot x-i\omega t}$,
we get from equation~(\ref{eq:linearized-driftkinetic})
\begin{align}
-i\omega\hat{f}_{\omega k}+ik_{\parallel}v\hat{f}_{\omega k}+\frac{1}{M_{e}}\left[ik_{\parallel}\hat{\phi}_{\omega k}\partial_{v}f_{eq}-i\omega\hat{A}_{\omega k}\partial_{v}f_{eq}\right] & =0\label{eq:fourier1}\\
k_{\perp}^{2}\hat{\phi}_{\omega k} & =C_{P}\int\hat{f}_{\omega k}\,dv,\label{eq:fourier2}\\
k_{\perp}^{2}\hat{A}_{\omega k} & =C_{A}\int v\hat{f}_{\omega k}\,dv,\label{eq:fourier3}
\end{align}
where $k_{\perp}^{2}=k_{1}^{2}+k_{2}^{2}$ and $k_{\parallel}=k_{3}$.
Since in this linear theory all the Fourier modes decouple we, from
now on, will suppress the indices $\omega$ and $k$. 

Plugging equations (\ref{eq:fourier2}) and (\ref{eq:fourier3}) into
equation (\ref{eq:fourier1}) we get
\[
i\left(-\omega+k_{\parallel}v\right)\hat{f}+\frac{ik_{\parallel}C_{P}}{k_{\perp}^{2}M_{e}}(\partial_{v}f_{eq})\left(\int\hat{f}\,dv\right)-\frac{i\omega C_{A}}{k_{\perp}^{2}M_{e}}(\partial_{v}f_{eq})\left(\int v\hat{f}\,dv\right)=0
\]
and thus
\[
\hat{f}-\frac{C_{P}}{M_{e}}\frac{k_{\parallel}}{k_{\perp}^{2}}\frac{1}{k_{\parallel}v-\omega}(\partial_{v}f_{eq})\left(\int\hat{f}\,dv\right)+\frac{C_{A}}{M_{e}}\frac{\omega}{k_{\perp}^{2}}\frac{1}{k_{\parallel}v-\omega}(\partial_{v}f_{eq})\left(\int v\hat{f}\,dv\right)=0.
\]
We can now express $\int v\hat{f}\,dv$ using $\int\hat{f}\,dv$ by
using the continuity equation
\[
\partial_{t}\int f\,dv+\partial_{z}\int vf\,dv=0.
\]
In frequency space we have 
\begin{align*}
-i\omega\int\hat{f}\,dv+ik\int v\hat{f}\,dv & =0
\end{align*}
and thus
\[
\int v\hat{f}\,dv=\frac{\omega}{k_{\parallel}}\int\hat{f}\,dv.
\]

We then have
\[
\hat{f}-\left[\frac{C_{P}}{M_{e}}\frac{k_{\parallel}}{k_{\perp}^{2}}\frac{1}{k_{\parallel}v-\omega}+\frac{C_{A}}{M_{e}}\frac{\omega}{k_{\perp}^{2}}\frac{\omega/k_{\parallel}}{k_{\parallel}v-\omega}\right](\partial_{v}f_{eq})\left(\int\hat{f}\,dv\right)=0
\]

By integrating in $v$ we obtain
\[
\left[1-\left(\frac{C_{P}}{M_{e}}\frac{k_{\parallel}}{k_{\perp}^{2}}-\frac{C_{A}}{M_{e}}\frac{\omega^{2}}{k_{\perp}^{2}k_{\parallel}}\right)\int\frac{\partial_{v}f_{eq}}{k_{\parallel}v-\omega}\,dv\right]\int\hat{f}\,dv=0
\]

from which we get
\[
1-\frac{1-\beta\omega^{2}/k_{\parallel}^{2}}{M_{e}\left(k_{\perp}\overline{\rho}_{i}\right)^{2}}\int\frac{\partial_{v}f_{eq}}{v-\omega/k_{\parallel}}\,dv=0,
\]
where $\overline{\rho}_{i}=\rho_{i}/L$. 

In order to being able to evaluate the integral we have to specify
$f_{eq}$. We will use a Maxwellian here, i.e.
\[
f_{eq}=\frac{\exp\left(-v^{2}/v_{th,e}^{2}\right)}{\sqrt{\pi}v_{th,e}}=\sqrt{\frac{M_{e}}{\pi}}\exp\left(-M_{e}v^{2}\right),
\]
since $v_{th,e}=\sqrt{2T_{\text{ref}}/m_{e}}=v_{th,i}/\sqrt{M_{e}}=1/\sqrt{M_{e}}$
(remember that we express velocities in units of $v_{th,i}$). 

It is useful to introduce $\overline{\omega}=\omega/(v_{th,e}k_{\parallel})=\sqrt{M_{e}}\omega/k_{\parallel}$.
Then

\begin{align*}
\int\frac{\partial_{v}f_{eq}}{v-\omega/k_{\parallel}}\,dv & =-M_{e}\int\frac{2v\sqrt{M_{e}}f_{eq}(v)}{v\sqrt{M_{e}}-\sqrt{M_{e}}\omega/k_{\parallel}}\,dv.
\end{align*}
Substituting $u=v\sqrt{M_{e}}$ we get
\begin{align*}
\int\frac{\partial_{v}f_{eq}}{v-\omega/k}\,dv & =-M_{e}\int\frac{2uf_{eq}(u/\sqrt{M_{e}})}{u-\overline{\omega}}\,du\\
 & =-2M_{e}\left[1+\overline{\omega}Z(\overline{\omega})\right],
\end{align*}
where

\[
Z(a)=\frac{1}{\sqrt{\pi}}\int\frac{e^{-v^{2}}}{v-a}\,dv=i\sqrt{\pi}e^{-a^{2}}\left(1+\text{erf}\left(ia\right)\right).
\]

We thus have
\[
1-\frac{2[1+\overline{\omega}Z(\overline{\omega})]}{\left(k_{\perp}\overline{\rho}_{i}\right)^{2}}(\beta\omega^{2}/k_{\parallel}^{2}-1)=0
\]
and therefore we get the dispersion relation
\[
1-\frac{2[1+\overline{\omega}Z(\overline{\omega})]}{\left(k_{\perp}\overline{\rho}_{i}\right)^{2}}(\beta/M_{e}\overline{\omega}^{2}-1)=0
\]

The two dimensionless parameters that determine $\overline{\omega}$
here are $\beta/M_{e}$ and $k_{\perp}\overline{\rho}_{i}$.

\end{document}